\documentclass[journal]{IEEEtran}
\usepackage[T1]{fontenc}% optional T1 font encoding

\usepackage{amssymb}
\usepackage{amsmath}
\usepackage{cite}
\usepackage{url}
\usepackage{cite,graphicx,amsmath,amssymb}
\usepackage{fancyhdr}
\usepackage{mdwmath}
\usepackage{mdwtab}
\usepackage{caption}
\usepackage{amsthm}
\usepackage{setspace}
\usepackage{hyperref}
\usepackage{algorithm}
\usepackage{algorithmic}
\usepackage{multirow}
\usepackage{makecell}
\usepackage{mathtools}
\usepackage{subcaption}
\usepackage{bm}
\usepackage{tikz}
\usepackage{xurl}
\usepackage{stfloats}
\usepackage{mathrsfs}

\usepackage{booktabs}
\usepackage{multirow}
\usepackage{siunitx}
\usepackage{balance}
\usepackage{bm}

\usepackage{array}
\usepackage{multirow,tabularx}
\usepackage{varwidth}

\hypersetup{colorlinks=true,
	linkcolor=blue,
	citecolor=blue,      
	urlcolor=black,
}

\expandafter\def\expandafter\normalsize\expandafter{%
	\normalsize%
	\setlength\abovedisplayskip{5.2pt}%
	\setlength\belowdisplayskip{5.2pt}%
	\setlength\abovedisplayshortskip{3pt}%
	\setlength\belowdisplayshortskip{3pt}%
}

\theoremstyle{definition}

\newtheorem{lemma}{Lemma}
\newtheorem{remark}{Remark}

\graphicspath{{./figure/}}

\begin{document}
	
	\title{Center-Fed Pinching Antenna System (C-PASS) Aided Wireless Communications}

	\author{Xu Gan,~\IEEEmembership{Member, IEEE}, and Yuanwei Liu,~\IEEEmembership{Fellow,~IEEE}
		\thanks{The authors are with the Department of Electrical and Electronic Engineering, The University of Hong Kong, Hong Kong (e-mail: \{eee.ganxu, yuanwei\}@hku.hk).}
	}
	
	\maketitle
	
	\begin{abstract}
		The novel architecture of the center-fed pinching antenna system (C-PASS) is investigated, where the waveguide-fed signal is divided into two propagation directions through controllable power splitting. By doing so, a doubled degree of freedom (DoF) is achieved compared to conventional PASS. Based on the new designed basic signal model of C-PASS, three practical operating protocols for C-PASS are proposed, namely power splitting (PS), direction switching (DS), and time switching (TS). Then, the sum-rate maximization problem for the joint optimization of transmit and pinching beamforming is formulated for each of the proposed protocols. 1) For PS, the highly coupled non-convex problem is first transformed into a tractable form via the weighted minimum mean square error reformulation and solved using the alternating optimization framework; 2) For DS, the above approach is subsequently extended to solve the mixed-integer constraints inherent for DS via the penalty-based algorithm; 3) For TS, the optimization problem can be decomposed into two subproblems and solved using the similar iterative techniques, while its optimal time allocation ratio is derived in closed form. Finally, numerical results reveal that TS is superior in the low-power regime, while PS and DS achieve significantly higher rates in the high-power regime due to the enhanced DoF.
		
		
	\end{abstract}

	\section{Introduction}
	\IEEEPARstart{T}{he} evolution towards beyond sixth-generation (B6G) wireless networks is envisioned to satisfy rigorous requirements for ultra-high spectral efficiency, ultra-dense connectivity, and energy sustainability\cite{10054381,akyildiz20206g}. To satisfy these ambitious goals, ``smart radio environment'' has attracted increasing attention by intelligently manipulating the electromagnetic propagation through \emph{programmable surfaces} \cite{gan2025multi} and \emph{advanced antenna architectures}\cite{han2023toward,new2023fluid}. Among them, the pinching antenna system (PASS)\cite{liu2025pinching,liu2025pinching2,yang2025pinching} has emerged as a promising architecture that enables highly flexible spatial reconfiguration. Specifically, signals are first fed into a dielectric waveguide and then efficiently radiated to serve users via dielectric pinching antennas (PAs) in PASS~\cite{wang2025modeling}. Through this distinctive architecture, PASS effectively mitigates large-scale path loss by substituting high-attenuation wireless links with waveguide transmission, while simultaneously exploiting small-scale fading via the precise tuning of PA radiation characteristics. Accordingly, PASS achieves significant performance gains via pinching beamforming by optimizing the radiation coefficients and positions of the PAs, in addition to the optimization of transmit beamforming through the baseband precoding~\cite{wang2025modeling,xu2025pinching,wang2025joint,zhao2025tri}.

	\subsection{Prior Works}
	Motivated by the favorable characteristics of PASS, extensive research efforts have been devoted to the utilizing of PASS aided wireless communications\cite{ding2025flexible,ding2025analytical,cheng2025performance,ouyang2025capacity,wang2025antenna,xie2025low,oikonomou2025ofdma,hou2025performance,shan2025exploiting,11050939,wang2025pinching,qian2025pinching,bereyhi2025mimo,zhao2025pinching,zhao2025waveguide,ouyang2025uplink}. In particular, existing studies on multi-user PASS communications are reviewed for \emph{single-waveguide} and \emph{multi-waveguide} architectures.

	\subsubsection{Studies on Single-Waveguide Aided Multi-User Communications}
	For the simplified two-user scenarios, the authors of \cite{ding2025flexible} have recognized that the single waveguide is fed with the same signal, inherently restricting spatial multiplexing. Consequently, they adopted non-orthogonal multiple access (NOMA) to leverage the additional degree of freedom (DoF) in the power domain for simultaneously serving two users. Specifically, \cite{ding2025flexible} derived the closed-form approximations for the ergodic sum rate under the NOMA scheme, accounting for the stochastic distribution of users. Motivated by these findings, the works in \cite{ding2025analytical} and \cite{cheng2025performance} conducted analytical investigations employing both orthogonal multiple access (OMA) and NOMA schemes for two-user scenarios. In particular, \cite{ding2025analytical} derived the optimal antenna location solutions to reveal the impact of PA placement on PASS performance, while \cite{cheng2025performance} derived closed-form expressions for the outage probabilities of the two users. To provide further theoretical insights into single-waveguide PASS, the authors of \cite{ouyang2025capacity} characterized the capacity regions under time-division multiple access (TDMA) and frequency-division multiple access (FDMA). Extending performance optimization to multi-user scenarios, the authors of \cite{wang2025antenna} and \cite{xie2025low} also employed NOMA strategies to maximize the system sum rate, for which low-complexity algorithms were developed to optimize PA placement and activation, respectively. For OMA schemes, the authors of \cite{oikonomou2025ofdma} and~\cite{hou2025performance} investigated the optimization designs of multi-user PASS enabled by OFDMA and TDMA, respectively. Furthermore, multicast communication proved well-suited to the single-waveguide constraint where identical signals are broadcast. Accordingly, the authors of~\cite{shan2025exploiting} and \cite{11050939} focused on PASS-aided multicast transmission, demonstrating substantial rate enhancements compared to conventional fixed-location antenna systems.

	\subsubsection{Studies on Multi-Waveguide Aided Multi-User Communications}
	Another intuitive approach to support multi-user transmission is to employ multiple waveguides, since different waveguides can transmit separate signals. In this case, an $M$-waveguide PASS is able to simultaneously serve $M$ users as investigated in \cite{wang2025pinching} and \cite{qian2025pinching}. Specifically, the authors of~\cite{wang2025pinching} introduced a waveguide assignment indicator to map each waveguide to a specific user, maximizing the system sum rate through assignment optimization. Besides, the authors of \cite{qian2025pinching} derived spectral efficiency expressions to quantify the resulting beamforming and multiplexing gains. For scenarios where $M$ waveguides support $K$ users ($M \ge K$), the PASS architecture can effectively support uplink and downlink transmissions within a MIMO framework \cite{bereyhi2025mimo}. In this case, the authors of \cite{bereyhi2025mimo} designed an optimal hybrid beamforming strategy to solve the downlink weighted sum-rate maximization and uplink multi-user detection. To enable greater flexibility at the baseband beamforming, the authors of \cite{zhao2025pinching} proposed three architectures: waveguide multiplexing, waveguide division, and waveguide switching. Based on these architectures, joint baseband signal processing and pinching beamforming designs were investigated for general multi-group multicast systems~\cite{zhao2025pinching}. More recently, the concepts of waveguide division multiple access (WDMA) \cite{zhao2025waveguide} and segmented waveguide-enabled pinching-antenna system (SWAN)~\cite{ouyang2025uplink} have been proposed for multi-user communications. More particularly, WDMA allocates a dedicated waveguide to each user, whereas SWAN utilizes a structure composed of multiple short waveguide segments to radiate or receive signals. In essence, both frameworks facilitate multi-user support through waveguide-level signal separation. For instance, the authors of~\cite{zhao2025waveguide} formulated the multi-waveguide power allocation problem to maximize the system sum rate. Furthermore, the authors of~\cite{ouyang2025uplink} enabled uplink SWAN communications by activating a single PA per segment, which could eliminate the inter-antenna radiation effect.

	\subsection{Motivations and Contributions}
	Most of the existing research contributions for PASS-aided multi-user communications can be broadly summarized along two directions. For the single-waveguide setting, multi-user access is typically supported by exploiting additional DoFs from extra resources~\cite{ding2025flexible,ding2025analytical,cheng2025performance,wang2025antenna,xie2025low,ouyang2025capacity,oikonomou2025ofdma,hou2025performance}, such as power-domain NOMA or time/frequency-domain OMA. In contrast, multi-waveguide architectures address this by exploiting the spatial DoFs for multi-stream transmissions across different waveguides~\cite{wang2025pinching,qian2025pinching,bereyhi2025mimo,zhao2025waveguide,ouyang2025uplink,zhao2025pinching}. However, neither strategy offers a sustainable path for B6G, as scaling either the additional resources or the waveguide number to accommodate ultra-dense connectivity is practically infeasible. To overcome the $\text{DoF}=1$ limitation in conventional PASS, a novel architecture of center-fed PASS (C-PASS) was proposed in \cite{gan2025c}. In particular, as illustrated in Fig.~\ref{fig:C-PASS}(a), the signal is fed into the center port of the waveguide and divided into two directions, denoted as forward-propagation (FP) and backward-propagation (BP). According to the theoretical results in \cite{gan2025c}, C-PASS doubles the available DoFs compared to conventional PASS architectures, thereby achieving significant spatial multiplexing gain. Despite these advantages, the investigation of how C-PASS can be utilized for wireless communications is still in its infancy. More particularly, efficient operating protocols and corresponding joint transmit and pinching beamforming optimization techniques for C-PASS aided wireless networks have not yet been studied, which constitutes the main motivation for this work.
	
	To fully exploit the potential of C-PASS, in this paper, we propose practical protocols and the joint transmit and pinching beamforming optimization algorithms for the operation of C-PASSs. The main contributions are summarized as follows:
	\begin{itemize}
		
		\item We propose three practical operating protocols for C-PASSs, namely power splitting (PS), direction switching (DS), and time switching (TS). Based on the derived in-waveguide channel expressions, we mathematically characterize these protocols for their optimization variables and distinct advantages.

		\item We consider a C-PASS aided downlink communications, where one center-fed waveguide serves multiple users by exploiting both FP and BP links. In particular, we formulate a joint transmit and pinching beamforming optimization problem for each of the proposed protocols to maximize the system sum rate. Specifically, the PS and DS protocols leverage simultaneous bidirectional propagation, where the power splitting ratios are modeled as continuous variables in $[0,1]$ and binary indicators in $\{0,1\}$, respectively. In contrast, under the TS protocol, the FP and BP links are activated in different periods, where the time allocation ratio is optimized to maximize the average system performance.

		\item For PS, we first transform the highly coupled non-convex optimization problem into a tractable form via the weighted minimum mean square error (WMMSE) reformulation. We then develop an efficient alternating optimization algorithm to iteratively update the transmit precoders, PA radiation coefficients, PA positions, and power splitting ratios. Furthermore, we extend this framework by integrating a penalty-based method to solve the mixed-integer constraints for DS. For TS, we decompose its optimization problem into two independent subproblems that are solved via similar iterative techniques, while the optimal time allocation ratio is derived in closed form.

		\item Finally, we provide extensive numerical results to validate the advantages of the proposed C-PASS architectures. Specifically, numerical results demonstrate that: 1) TS is superior in the low-power regime, whereas PS and DS double the sum rate in the high-SNR regime due to the enhanced DoF; 2) PS and DS yield higher transmit beamforming gains that increase with the number of input ports compared to conventional PASS; 3) The pinching beamforming gain is more significant under PS and DS than under TS, and this advantage becomes more pronounced with the number of PAs.

	\end{itemize}

	\subsection{Organization and Notation}
	The remainder of this paper is structured as follows. Section~\ref{sec:model} provides the basic signal model and practical operating protocols for C-PASS. Section~\ref{sec:system} introduces the system model and formulates the sum-rate maximization problem for PS, DS, and TS protocols. Section~\ref{sec:solutions} proposes the joint transmit and pinching beamforming algorithms in an alternating optimization framework to provide C-PASS design solutions for the three protocols. Numerical results evaluate the sum-rate performance of the C-PASS under the three protocols and compare them with the conventional end-fed PASS in Section~\ref{sec:simulation}. Finally, Section \ref{sec:conclusion} concludes the paper.

	\emph{Notations:} Boldface upper-case letters, boldface lower-case letters, and normal lower-case letters describe matrices, vectors, and scalars, respectively. The spaces of real and complex numbers are symbolized by $\mathbb{R}$ and $\mathbb{C}$. The operators $(\cdot)^T$, $(\cdot)^H$, $(\cdot)^{-1}$, and $\text{Tr}(\cdot)$ denote the transpose, conjugate transpose (Hermitian), inverse, and trace, respectively. The notation $[\mathbf{A}]_{m,n}$ refers to the entry located at the $m$-th row and $n$-th column of $\mathbf{A}$, while $\mathbf{I}_N$ stands for the $N \times N$ identity matrix. We use $\text{blkdiag}\{\mathbf{x}_1, \dots, \mathbf{x}_M\}$ to represent a block diagonal matrix formed by the diagonal components $\mathbf{x}_1, \dots, \mathbf{x}_M$. Furthermore, $|\cdot|$ and $\|\cdot\|$ indicate the absolute value and the Euclidean norm, respectively. The Hadamard (element-wise) product is marked by $\odot$. Finally, $\mathbb{E}[\cdot]$ represents the statistical expectation, and $\Re \{\cdot\}$ extracts the real component of a complex value

	\section{Basic Signal Model and Practical Operating Protocols for C-PASS}\label{sec:model}
	In this section, we present the basic signal model and practical operating protocols for C-PASS aided wireless communications.
	
	\subsection{Basic Signal Model of C-PASS}
	\begin{figure*}[t]
		\centering
		\includegraphics[width=1\linewidth]{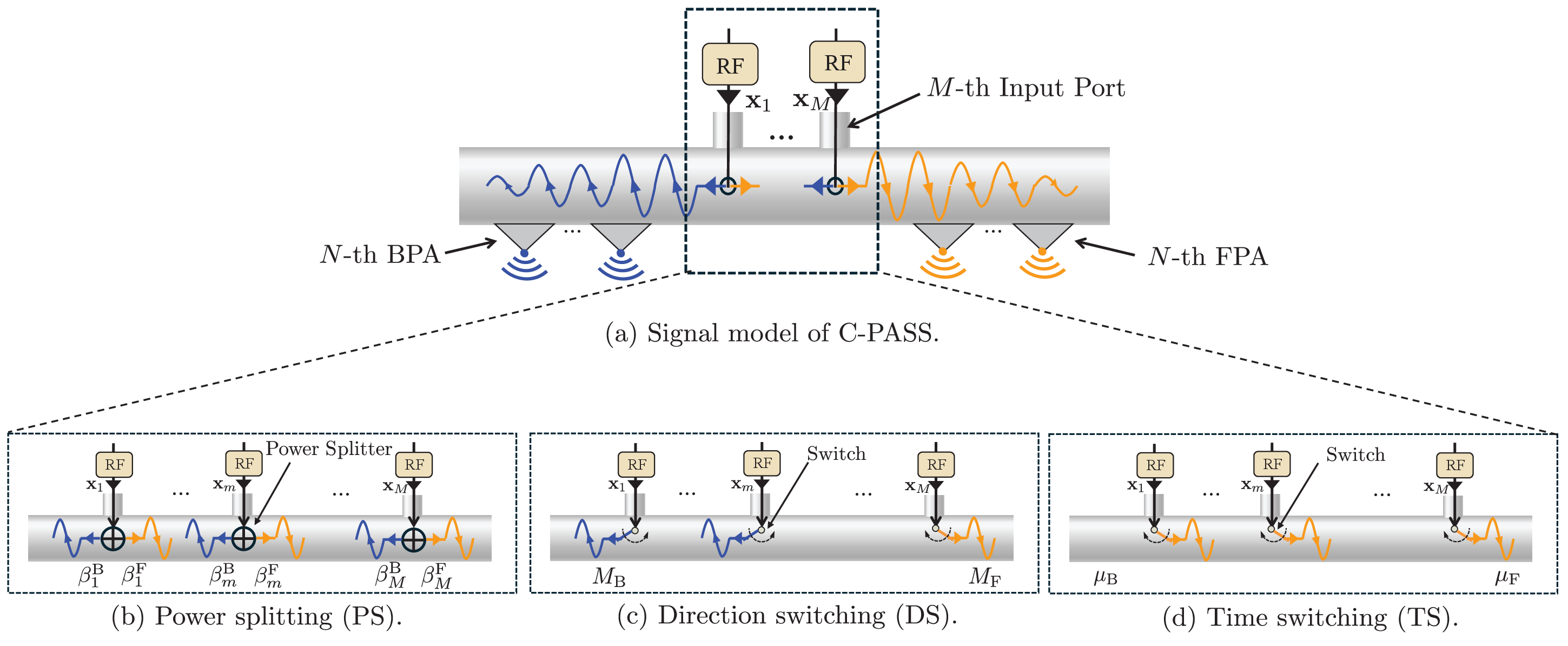}
		\caption{Three practical protocols for operating C-PASS.}
		\label{fig:C-PASS}
	\end{figure*}
	
	To characterize the bidirectional propagation model of the C-PASS architecture~\cite{gan2025c} shown in Fig.~\ref{fig:C-PASS}(a), let $\mathbf{x}_m$ denote the signal fed into the $m$-th input port. The signal is subsequently divided into the FP component $\sqrt{\beta_m^{\text{F}}} \mathbf{x}_m$ and the BP component $\sqrt{\beta_m^{\text{B}}} \mathbf{x}_m$ at the tunable power splitter\cite{reichel2016broadband}. Assuming the operation of the $M$ power splitters incurs no power loss, the law of conservation of energy for $\forall m$ yields
	\begin{equation}\label{sec2_s1}
		\beta_m^{\text{F}} + \beta_m^{\text{B}} = 1.
	\end{equation} 
	These split signals are then radiated into free space via the forward-propagation PAs (FPAs) and backward-propagation PAs (BPAs). Based on the PA radiation model in \cite{gan2025dual}, the radiated signal expression at the $n$-th $\chi$-PA is given by
	\begin{equation}\label{sec2_PA}
		\mathbf{x}_{n}^{\chi} = \sum_{m=1}^M \sqrt{\beta_m^{\chi} \xi_n^{\chi}} \exp\left( -j k_g D_n^{\chi} \right) \mathbf{x}_m,
	\end{equation}
	where $\chi \in \{ \text{F}, \text{B} \}$ denotes the FP or BP direction. The terms $\xi_n^{\chi}$ and $D_n^{\chi}$ represent the cumulative radiation coefficient and propagation distance for the $n$-th $\chi$-PA, respectively. $k_g=2\pi/ \lambda_g$ is the propagation wavenumber in the waveguide, and $\lambda_g$ is its effective wavelength. Particularly, the in-waveguide $\lambda_g$ is related to the free-space $\lambda_0$ by the effective refractive index $n_{\text{eff}}$ of the waveguide, denoting $\lambda_g=\lambda_0/n_{\text{eff}}$.

	\subsection{Three Practical Protocols for Operating C-PASS}
	As can be observed from \eqref{sec2_PA}, by properly adjusting the power splitting ratios $\{\beta_m^{\chi}\}_{m=1}^M$, the signal fed into the $m$-th input port can be operated in the full FP direction (i.e., $\beta_m^{\text{F}}=1$), the full BP direction (i.e., $\beta_m^{\text{B}}=1$), and the general simultaneous FP and BP directions (i.e., $\beta_m^{\text{F}}, \beta_m^{\text{B}} \in [0,1]$). Inspired by these observations, we propose three practical protocols for operating C-PASS with $M$ input ports, namely power splitting (PS), direction switching (DS), and time switching (TS), as illustrated in Fig.~\ref{fig:C-PASS}(b)-(d). To mathematically characterize the affect of these protocols on the in-waveguide channels, let $\mathbf{G}_{\chi}^{\Psi} \in \mathbb{C}^{M \times N}$ denote the channel from the input ports to the $\chi$-PAs, for the $\Psi \in \{\text{PS}, \text{DS}, \text{TS} \}$ protocol in the $\chi \in \{ \text{F}, \text{B} \}$ direction.
	
	\subsubsection{Power Splitting}
	For PS, shown in Fig.~\ref{fig:C-PASS}(b), all input signals are splitted into both FP and BP directions with the power splitting ratio $\beta_m^{\text{F}}$ and $\beta_m^{\text{B}}$, respectively, for $\forall m$. In this case, the channel coefficient from the $m$-th input port to the $n$-th $\chi$-PA is given by
	\begin{subequations}
		\begin{align}
			& \left[ \mathbf{G}_{\text{F}}^{\text{PS}} \right]_{m,n}=\sqrt{\beta_m^{\text{F}}} \exp\left(-j k_g  |\mathbf{p}_n^{\text{FPA}}-\mathbf{p}_m^{\text{IN}} |  \right), \\
			& \left[ \mathbf{G}_{\text{B}}^{\text{PS}} \right]_{m,n}=\sqrt{\beta_m^{\text{B}}} \exp\left(-j k_g  |\mathbf{p}_n^{\text{BPA}}-\mathbf{p}_m^{\text{IN}} |  \right), 
		\end{align}
	\end{subequations}
	where $\mathbf{p}_n^{\chi\text{PA}}$ and $\mathbf{p}_m^{\text{IN}}$ denotes the position coordinates of the $n$-th $\chi$-PA and $m$-th input port, respectively. The power splitting ratio satisfies $\beta_m^{\text{F}}, \beta_m^{\text{B}} \in [0,1]$ and $\beta_m^{\text{F}} + \beta_m^{\text{B}} = 1$. Among these three protocols, the PS offers the highest degree of design flexibility by allowing for the simultaneous optimization of bidirectional power splitting ratios at each input port. However, this advantage comes at the expense of increased computational complexity and substantial hardware costs required for precise control.
	
	\subsubsection{Direction Switching}
	For DS, as shown in Fig.~\ref{fig:C-PASS}(c), all input ports can be divided into two groups. Specifically, one group contains $M_{\text{F}}$ input ports that operated in the full FP direction, while the other group contains $M_{\text{B}}$ input ports operating in the full BP direction, where $M_{\text{F}}+M_{\text{B}}=M$. Accordingly, the channel coefficient from the $m$-th input port to the $n$-th $\chi$-PA is given by
	\begin{subequations}
		\begin{align}
			& \left[ \mathbf{G}_{\text{F}}^{\text{DS}} \right]_{m,n}=\sqrt{\beta_m^{\text{F}}} \exp\left(-j k_g  |\mathbf{p}_n^{\text{FPA}}-\mathbf{p}_m^{\text{IN}} |  \right), \\
			& \left[ \mathbf{G}_{\text{B}}^{\text{DS}} \right]_{m,n}=\sqrt{\beta_m^{\text{B}}} \exp\left(-j k_g  |\mathbf{p}_n^{\text{BPA}}-\mathbf{p}_m^{\text{IN}} |  \right), 
		\end{align}
	\end{subequations}
	where $\beta_m^{\text{F}}, \beta_m^{\text{B}} \in \{0,1\}$ and $\beta_m^{\text{F}} + \beta_m^{\text{B}} = 1$. We note that DS can be regarded as a special case of PS, where the power splitting ratios for FP and BP are restricted to binary values. In this case, the DS protocol can adopt the ``on-off'' switching, which is much easier to implement compared to PS.

	\subsubsection{Time Switching}  
	Different from PS and DS, the TS periodically switches all input ports in the full FP or BP direction for different periods, as illustrated in Fig.~\ref{fig:C-PASS}(d). Let $0 \le \mu_{\text{FU}} \le 1$ and $0 \le \mu_{\text{BU}} \le 1$ denote the time allocation ratio for the FP and BP, respectively, with $\mu_{\text{FU}} + \mu_{\text{BU}} = 1$. Then, the channel coefficient can be written as
	\begin{subequations}
		\begin{align}
			& \left[ \mathbf{G}_{\text{F}}^{\text{TS}} \right]_{m,n}=\exp\left(-j k_g  |\mathbf{p}_n^{\text{FPA}}-\mathbf{p}_m^{\text{IN}} |  \right), \\
			& \left[ \mathbf{G}_{\text{B}}^{\text{TS}} \right]_{m,n}=\exp\left(-j k_g  |\mathbf{p}_n^{\text{BPA}}-\mathbf{p}_m^{\text{IN}} |  \right),
		\end{align}
	\end{subequations}
	respectively for $\mu_{\text{FU}}$ and $\mu_{\text{BU}}$ period. Different from PS and DS, the design of the FP- and BP-direction communications for TS is not coupled due to the exploitation of the orthogonal time resources, and thus easier to handle.

	In Table~\ref{tab:protocols}, we summarize the optimization variables and advantages for the PS, DS, and TS operating protocols.
	
	\begin{table*}[t]
		\centering
		\caption{Summary of optimization variables and advantages for these three protocols.}
		\label{tab:protocols}
		
		\begin{tabular}{|>{\centering\arraybackslash}m{1.6cm}|
				>{\centering\arraybackslash}m{4.5cm}|
				>{\centering\arraybackslash}m{4.7cm}|
				>{\centering\arraybackslash}m{5cm}|}
			\hline
			\textbf{Protocols} & \textbf{Power Splitting Ratios} & \textbf{Time Allocation Ratios} & \textbf{Advantages} \\
			\hline
			PS &
			$\beta_m^{\text{F}}, \beta_m^{\text{B}}\in[0,1], \beta_m^{\text{F}}+\beta_m^{\text{B}}=1$ &
			$\times$ &
			High flexibility \\
			\hline
			DS &
			$\beta_m^{\text{F}}, \beta_m^{\text{B}}\in\{0,1\}, \beta_m^{\text{F}}+\beta_m^{\text{B}}=1$ &
			$\times$ &
			Easy implementation \\
			\hline
			TS &
			$\times$ &
			$\mu_{\text{FU}}, \mu_{\text{BU}}\in[0,1], \mu_{\text{FU}}+\mu_{\text{BU}}=1$ &
			Decoupled optimization for FP/BP \\
			\hline
		\end{tabular}
	\end{table*}

	\section{System Model and Problem Formulation}\label{sec:system}
	Based on the proposed protocols, in this section, we present the system model of C-PASS aided downlink communications and formulate joint transmit and pinching beamforming optimization problems to maximize system sum-rate.
	
	\subsection{System Model}
	\begin{figure}[t]
		\centering
		\includegraphics[width=1\linewidth]{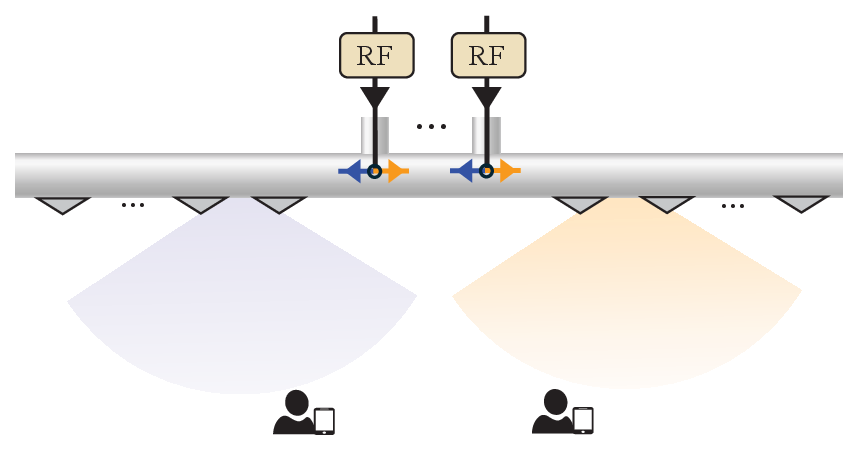}
		\caption{Illustration of a C-PASS aid downlink communications.}
		\label{fig:system_model}
	\end{figure}
	
	As shown in Fig.~\ref{fig:system_model}, we consider a C-PASS aided downlink communications, where one waveguide is center fed through $M$ input ports and then radiated by $N$ FPAs and $N$ BPAs. Without loss of generality, taking the waveguide center as the origin, $M$ input ports are uniformly and symmetrically distributed over the origin at the interval length of $L_{\text{IN}}$. As such, the position coordinates of the $m$-th input port is written as $\mathbf{p}_m^{\text{IN}} = \left( (m-\frac{M+1}{2})L_{\text{IN}}, 0 \right)$. For simplicity of presentation, we consider two users located close to the FPA and BPA, denoted as the FP-direction user (FU) and BP-direction user (BU), respectively. The position coordinates of FU and BU are given by $\mathbf{p}_{\text{FU}}=\left[ X_{\text{FU}}, Y_{\text{FU}}  \right]$ and  $\mathbf{p}_{\text{BU}}=\left[ X_{\text{BU}}, Y_{\text{BU}}  \right]$, respectively. Let $\mathbf{h}_k^{\text{F}} \in \mathbb{C}^{N \times 1}$ and $\mathbf{h}_k^{\text{B}} \in \mathbb{C}^{N \times 1}$ denote the channel from the FPA and BPA to the $k$-user, respectively, where $k \in \{\text{FU}, \text{BU} \}$. Specifically, the free-space propagation channel can be modeled by
	\begin{equation}
		\left[\mathbf{h}_k^{\text{F}} \right]_n = \eta \frac{e^{-j k_0 r_{k,n}^{\text{F}} }}{r_{k,n}^{\text{F}}}, \quad \left[\mathbf{h}_k^{\text{B}} \right]_n = \eta \frac{e^{-j k_0 r_{k,n}^{\text{B}} }}{r_{k,n}^{\text{B}}},
	\end{equation}
	where $\eta$ is the effective channel factor accounting for the free-space path loss. $r_{k,n}^{\text{F}}$ and $r_{k,n}^{\text{B}}$ represents the distance from the $n$-th FPA and BPA to the $k$-user, respectively, i.e., $r_{k,n}^{\text{F}} = |\mathbf{p}_n^{\text{FPA}} - \mathbf{p}_k| $ and $r_{k,n}^{\text{B}} = |\mathbf{p}_n^{\text{BPA}} - \mathbf{p}_k| $. Based on the considered system model, the received signal at the $k$-user under the $\Psi$-protocol is
	\begin{align}
		y_k^{\Psi} = \sum_{\chi \in \{\text{F}, \text{B} \}} (\mathbf{h}_k^{\chi})^T\bm{\Sigma}_{\chi} (\mathbf{G}_{\chi}^{\Psi})^T \mathbf{W} \mathbf{s} + n_k,
	\end{align}
	with $\Psi \in \{ \text{PS},\text{DS}, \text{TS} \}$. $\bm{\Xi}_{\text{F}} = \text{diag}\{\sqrt{\xi_1^{\text{F}}},\sqrt{\xi_2^{\text{F}}},\ldots,\sqrt{\xi_N^{\text{F}}} \}$ and $\bm{\Xi}_{\text{B}}=\text{diag}\{ \sqrt{\xi_1^{\text{B}}},\sqrt{\xi_2^{\text{B}}},\ldots,\sqrt{\xi_N ^{\text{B}}}\}$ denote the power radiation matrix of the FPAs and BPAs, respectively. $\mathbf{W} = \left[ \mathbf{w}_{\text{FU}} , \mathbf{w}_{\text{BU}} \right]$ is the BS precoding matrix, satisfying $\sum_{k \in \{ \text{FU},\text{BU} \} } \|\mathbf{w}_k\|^2 \le P_T$ with $P_T$ being the maximum transmit power. $\mathbf{s} = [s_{\text{FU}}, s_{\text{BU}}]$ is the information-bearing symbol for the FU and BU, with $\mathbb{E}\{|s_k|^2\}=1$. $n_k \sim \mathcal{CN}(0,N_0)$ denotes the additive noise with the power of $N_0$.

	\subsection{Problem Formulation}
	We derive the achievable sum-rate expressions under the PS, DS, and TS protocols, and formulate the corresponding sum-rate maximization problems.
	
	\subsubsection{PS and DS}
	The achievable rate of the user $k$ for PS or DS is given by
	\begin{equation}\label{sec3_R_ES}
		R_k^{\text{PS/DS}} \!=\! \log_2 \!\!\Bigg(\!\! 1+ \frac{\left| \sum_{\chi \in \{\text{F}, \text{B} \}} (\mathbf{h}_k^{\chi})^T\bm{\Sigma}_{\chi}\! (\mathbf{G}_{\chi}^{\text{PS/DS}})^T  \mathbf{w}_k \right|^2}{\left|  \sum_{\chi \in \{\text{F}, \text{B} \}} (\mathbf{h}_k^{\chi})^T\bm{\Sigma}_{\chi}\! (\mathbf{G}_{\chi}^{\text{PS/DS}})^T \mathbf{w}_{\bar{k}}   \right|^2\!\! + \! N_0}  \Bigg),
	\end{equation}
	where $\bar{k}$ denotes the other element of $\{\text{FU}, \text{BU}\}$ distinct from $k$. To maximize the achievable sum rate~\eqref{sec3_R_ES}, the optimization design involves the \emph{transmit beamforming} characterized by the power splitting ratios $\bm{\beta}_{\chi}$ and the precoding matrix $\mathbf{W}$, and the \emph{pinching beamforming} characterized by the PA radiation matrix $\bm{\Sigma}_{\chi}$ and the position coordinates $\mathbf{P}_{\text{PA}}=[\mathbf{p}_1^{ \text{FPA}},\dots,\mathbf{p}_N^{ \text{FPA}}, \mathbf{p}_1^{ \text{BPA}},\dots,\mathbf{p}_N^{ \text{BPA}}]$. Then, the optimization problem can be formulated as follows:
	\begin{subequations}\label{sec3_PSDS_problem}
		\begin{align}
			\max_{\bm{\beta}_{\chi},\mathbf{W},\bm{\Sigma}_{\chi},\mathbf{P}_{\text{PA}}}
			\ & R_{\text{FU}}^{\text{PS/DS}} +  R_{\text{BU}}^{\text{PS/DS}} \label{sec3_PSDS_problem_function}\\
			\text{s.t.}\quad
			& \beta_m^{\text{F}} + \beta_m^{\text{B}} = 1, \forall m,\\
			& \beta_m^{\text{F}}, \beta_m^{\text{B}} \in [0,1], \forall m, \text{for PS}, \\
			& \beta_m^{\text{F}}, \beta_m^{\text{B}} \in \{0,1\}, \forall m, \text{for DS}, \\
			& \|\mathbf{W}\|^2 \le P_T \\
			& \|\bm{\Sigma}_{\chi}\|^2 = 1, \forall \chi \\
			& \mathbf{P}_{\text{PA}} \in \mathcal{G}(\mathbf{P}_{\text{PA}}),
		\end{align}
	\end{subequations}
	where $\mathcal{G}(\mathbf{P}_{\text{PA}})$ is the feasible set of PA deployment.
	
	\subsubsection{TS}
	For TS, the BS consecutively sends $\mathbf{w}_{\text{FU}} s_{\text{FU}}$ and $\mathbf{w}_{\text{BU}} s_{\text{BU}}$ in the $\mu_{\text{FU}}$ and $\mu_{\text{BU}}$ period, respectively. Thus, the achievable communication rate of $k$-user can be expressed as
	\begin{equation}\label{sec3_R_TS}
		R_k^{\text{TS}} = \mu_k \log_2 \Bigg( 1+ \frac{\left| \sum_{\chi \in \{\text{F}, \text{B} \}} (\mathbf{h}_k^{\chi})^T\bm{\Sigma}_{\chi} (\mathbf{G}_{\chi}^{\text{TS}})^T  \mathbf{w}_k \right|^2}{\mu_k N_0} \Bigg),
	\end{equation}
	which represents the average communication rate accounting for the time allocation ratio $\bm{\mu}=[\mu_{\text{FU}}, \mu_{\text{BU}}]^T$. Then, the rate maximization problem is written as
	\begin{subequations}\label{sec3_TS_problem}
		\begin{align}
			\max_{\bm{\mu},\mathbf{W},\bm{\Sigma}_{\chi},\mathbf{P}_{\text{PA}}}
			\ & R_{\text{FU}}^{\text{TS}} +  R_{\text{BU}}^{\text{TS}} \label{sec3_TS_problem_function}\\
			\text{s.t.}\quad
			& \mu_{\text{FU}}+\mu_{\text{BU}}=1, \\
			& \mu_{\text{FU}},\mu_{\text{BU}} \in [0,1] ,\\
			& \|\mathbf{W}\|^2 \le P_T, \\
			& \|\bm{\Sigma}_{\chi}\|^2 = 1, \forall \chi, \\
			& \mathbf{P}_{\text{PA}} \in \mathcal{G}(\mathbf{P}_{\text{PA}}),
		\end{align}
	\end{subequations}
	
	\begin{remark}
		Note that the sum-rate maximization problem for the conventional end-fed PASS \cite{xu2025rate} can be regarded as a special case of the proposed C-PASS framework. Owing to the enhanced design DoF for power splitting or time allocation ratios, the C-PASS inherently offers superior flexibility and guarantees a sum-rate performance no worse than that of its end-fed counterpart. This enhanced performance comes at the expense of increased optimization challenges and computational complexity. In particular, the newly introduced power splitting ratio $\bm{\beta}_{\chi}$ for PS/DS or time allocation ratio $\bm{\mu}$ for TS is intrinsically coupled with the conventional beamforming variables $\mathbf{W}$, $\bm{\Sigma}_{\chi}$, and $\mathbf{P}_{\text{PA}}$. This motivates the design of optimization algorithms for three protocols to effectively solve these challenges and achieve sum-rate maximization.
	\end{remark}

	\section{Solutions for the three proposed Protocols}\label{sec:solutions}
	In this section, we first propose an efficient alternating optimization algorithm tailored for the PS protocol to solve the joint transmit and pinching beamforming design for sum-rate maximization in C-PASS. This algorithm is then extended to the DS protocol, where a penalty-based approach is introduced to handle the binary constraints. Furthermore, for the TS protocol, similar iterative techniques are employed, where the optimal time allocation ratio is derived in closed form.

	\subsection{The Proposed Solution for PS}
	To facilitate the subsequent algorithm derivations, we first define the effective channel from the input ports to the $k$-user as $\mathbf{h}_{k,\text{eff}} = \sum_{\chi \in \{\text{F}, \text{B} \}} \mathbf{G}_{\chi}^{\Psi} \bm{\Sigma}_{\chi} \mathbf{h}_k^{\chi}$. Then, the objective function~\eqref{sec3_PSDS_problem_function} can be rewritten as
	\begin{equation}\label{sec4_PS_o1}
		\max_{\bm{\beta}_{\chi},\mathbf{W},\bm{\Sigma}_{\chi},\mathbf{P}_{\text{PA}}} \sum_{k \in \{ \text{FU},\text{BU} \} } \log_2 \!\left(\! 1\!+\! \frac{\left|\mathbf{h}_{k,\text{eff}}^T  \mathbf{w}_k \right|^2}{\left|  \mathbf{h}_{k,\text{eff}}^T \mathbf{w}_{\bar{k}}   \right|^2\! + \! N_0}  \right).
	\end{equation}
	The above optimization problem remains highly coupled due to the logarithmic-fractional structure~\eqref{sec4_PS_o1}. To address this issue, we introduce the following lemma to transform the sum-rate optimization into a more tractable expression.
	
	\begin{lemma}
		The sum-rate maximization function~\eqref{sec4_PS_o1} for PS is equivalent to the weighted error minimization function as \cite{shi2011iteratively}
		\begin{equation}\label{sec4_PSDS_lemma1}
			\min_{\bm{\kappa},\mathbf{t},\bm{\beta}_{\chi},\mathbf{W},\bm{\Sigma}_{\chi},\mathbf{P}_{\text{PA}} } \ \sum_{k \in \{ \text{FU},\text{BU} \} } \kappa_{k} \epsilon_{k} - \ln(\kappa_{k}),
		\end{equation}
		where $\bm{\epsilon}=\left[\epsilon_{\text{FU}},\epsilon_{\text{BU}}\right]$, $\bm{\kappa}=\left[\kappa_{\text{FU}},\kappa_{\text{BU}}  \right]$ and $\mathbf{t} = [t_{\text{FU}}, t_{\text{BU}}]$ are the auxiliary vectors, and the error is defined as 
		\begin{align}
			&\epsilon_k = |t_k|^2 \left( \left|\mathbf{h}_{k,\text{eff}}^T \mathbf{w}_k\right|^2  +\left| \mathbf{h}_{k,\text{eff}}^T \mathbf{w}_{\bar{k}} \right|^2 + N_0   \right) \notag \\
			&\qquad - 2 \Re\left\{ t_k^* \mathbf{h}_{k,\text{eff}}^T \mathbf{w}_k \right\}  +1.
		\end{align}
		\begin{proof}
			Please refer to the proof derivation of [\emph{Theorem 1},~\citen{shi2011iteratively}].
		\end{proof}
	\end{lemma}
	
	Since the objective function~\eqref{sec4_PSDS_lemma1} is convex with respect to (w.r.t.) the auxiliary variables $\bm{\kappa}$ and $\mathbf{t}$, their optimal values can be obtained in closed form. By setting their first-order derivatives to zero, the optimal $\bm{\kappa}$ and $\mathbf{t}$ are derived as
	\begin{subequations}\label{update_t_kappa_PS}
		\begin{align}
			& t_k^{\text{opt}} = \frac{\mathbf{h}_{k,\text{eff}}^T \mathbf{w}_k}{ \left| \mathbf{h}_{k,\text{eff}}^T \mathbf{w}_k\right|^2  \!+ \! \left| \mathbf{h}_{k,\text{eff}}^T \mathbf{w}_{\bar{k}} \right|^2  + N_0}, \\
			&   \kappa_k^{\text{opt}} = \epsilon_k^{-1}.
		\end{align}
	\end{subequations}
	By substituting the derived optimal $\bm{\kappa}^{\text{opt}}$ and $\mathbf{t}^{\text{opt}}$ into the objective function~\eqref{sec4_PSDS_lemma1}, the optimization problem for PS reduces to
	\begin{subequations}\label{sec4_problem_PS_all}
		\begin{align}
			\min_{\bm{\beta}_{\chi},\mathbf{W},\bm{\Sigma}_{\chi},\mathbf{P}_{\text{PA}}} \quad
			& f_{\text{all}}^{\text{PS}}(\bm{\beta}_{\chi},\mathbf{W},\bm{\Sigma}_{\chi},\mathbf{P}_{\text{PA}}) \label{sec4_PS_all_ob} \\
			\text{s.t.}\quad
			& \beta_m^{\text{F}} + \beta_m^{\text{B}} = 1, \forall m,\\
			& \beta_m^{\text{F}}, \beta_m^{\text{B}} \in [0,1], \forall m,\label{sec4_PS_con} \\
			& \|\mathbf{W}\|^2 \le P_T, \\
			& \|\bm{\Sigma}_{\chi}\|^2 = 1, \forall \chi, \\
			& \mathbf{P}_{\text{PA}} \in \mathcal{G}(\mathbf{P}_{\text{PA}}),
		\end{align}
	\end{subequations}
	where the objective function~\eqref{sec4_PS_all_ob} is given by
	\begin{equation}
		\begin{aligned}
			f_{\text{all}}^{\text{PS}}&(\bm{\beta}_{\chi},\mathbf{W},\bm{\Sigma}_{\chi},\mathbf{P}_{\text{PA}}) =\sum_{k \in \{ \text{FU},\text{BU} \} } \kappa_k  \Big[ |t_k|^2\\
			&\times \left( \left|\mathbf{h}_{k,\text{eff}}^T \mathbf{w}_k \right|^2  + \left| \mathbf{h}_{k,\text{eff}}^T \mathbf{w}_{\bar{k}} \right|^2 \right)- 2 \Re\left\{ t_k^* \mathbf{h}_{k,\text{eff}}^T \mathbf{w}_{k} \right\} \Big].
		\end{aligned}
	\end{equation}
	Based on this reformulated formulation~\eqref{sec4_problem_PS_all}, we proceed to solve the optimization problem using an alternating iterative framework. Specifically, the variables $\mathbf{W}$, $\bm{\Sigma}_{\chi}$, $\mathbf{P}_{\text{PA}}$, and $\bm{\beta}_{\chi}$ are updated sequentially, while keeping the remaining variables fixed.
	
	\subsubsection{Subproblem w.r.t. $\mathbf{W}$}
	Fixing $\{ \bm{\beta}_{\chi}, \bm{\Sigma}_{\chi}, \mathbf{P}_{\text{PA}} \}$ and discarding terms constant with respect to $\mathbf{W}$ in \eqref{sec4_problem_PS_all}, the subproblem w.r.t. $\mathbf{W}$ is simplified as
	\begin{subequations}\label{sec4_PD_W}
		\begin{align}
			&\min_{\mathbf{W}} \
			f_{\mathbf{W}}^{\text{PS}} \label{sec4_ES_W_fun} \\
			&\text{s.t.}\quad
			\|\mathbf{w}_{\text{FU}}\|^2 + \|\mathbf{w}_{\text{BU}}\|^2 \le P_T, \label{sec4_ES_W_con}
		\end{align}
	\end{subequations}
	where the objective function is given by
	\begin{equation}
		\begin{aligned}
			f_{\mathbf{W}}^{\text{PS}} =&\ \mathbf{w}_{\text{FU}}^H  \sum_{k \in \{ \text{FU},\text{BU} \} }   \kappa_k  |t_k|^2 \mathbf{h}_{k,\text{eff}}^* \mathbf{h}_{k,\text{eff}}^T  \mathbf{w}_{\text{FU}} \\
			& + \mathbf{w}_{\text{BU}}^H  \sum_{k \in \{ \text{FU},\text{BU} \} }  \kappa_k  |t_k|^2 \mathbf{h}_{k,\text{eff}}^* \mathbf{h}_{k,\text{eff}}^T \mathbf{w}_{\text{BU}} \\
			& -  2\sum_{k \in \{ \text{FU},\text{BU} \} } \Re \left\{\kappa_k t_k^* \mathbf{h}_{k,\text{eff}}^T  \mathbf{w}_{k} \right\}.
		\end{aligned}
	\end{equation}
	To address the power constraint \eqref{sec4_ES_W_con}, we introduce a Lagrange multiplier $\lambda \ge 0$. The associated Lagrangian function is then constructed as
	\begin{equation}\label{sec4_ES_W_obj2}
		\min_{\mathbf{W}} \ f_{\mathbf{W}}^{\text{PS}}+ \lambda \left( \|\mathbf{w}_{\text{FU}}\|^2 + \|\mathbf{w}_{\text{BU}}\|^2 - P_T \right).
	\end{equation}
	Since \eqref{sec4_ES_W_obj2} is an unconstrained convex function w.r.t. $\mathbf{W}$, the optimal solution can be derived by setting its first-order derivative to zero. This yields the closed-form expression for the optimal precoding:
	\begin{equation}\label{update_W_PS}
		\begin{aligned}
			\mathbf{w}_k =  \! \Big[\! \sum_{i \in \{ \text{FU},\text{BU} \} } \!\! \kappa_{i} |t_i|^2 \mathbf{h}_{i,\text{eff}}^* \mathbf{h}_{i,\text{eff}}^T + \lambda \mathbf{I}_M \Big]^{-1}\!\! \kappa_k t_k \mathbf{h}_{k,\text{eff}}^*, \\
		\end{aligned}
	\end{equation}
	where the optimal $\lambda$ is chosen to satisfy the complementary slackness condition $\lambda^{\text{opt}} ( \sum_{i \in \{ \text{FU},\text{BU} \} } \|\mathbf{w}_i\|^2 - P_T ) = 0$.

	\subsubsection{Subproblem w.r.t. $\mathbf{\Xi}_{\chi}$}
	To facilitate the optimization of $\bm{\Sigma}_{\chi}$, we extract its diagonal elements and define the equivalent radiation vector $\bm{\xi}_{\chi}$, where $\bm{\Sigma}_{\chi} = \text{diag}\left(\bm{\xi}_{\chi}\right)$. Consequently, the subproblem of the PA power radiation is equivalently transformed into 
	\begin{subequations}\label{sec4_ES_xi}
		\begin{align}
			&\min_{\bm{\xi}_{\chi}} \
			f_{\bm{\xi}}^{\text{PS}} \label{sec4_ES_xi_fun} \\
			&\text{s.t.}\quad
			\left\| \bm{\xi}_{\text{F}} \right\|^2 =1, \ \left\| \bm{\xi}_{\text{B}} \right\|^2 =1,
		\end{align}
	\end{subequations}
	where the objective function is given by
	\begin{equation}
		\begin{aligned}
			& f_{\bm{\xi}}^{\text{PS}} = \left( \bm{\xi}_{\text{F}} \right)^T \Big[ \underbrace{ \sum_{k \in \{ \text{FU},\text{BU} \} } \kappa_k |t_k|^2  \mathbf{R}_k^{\text{F}} \mathbf{W}_{\text{sum}} (\mathbf{R}_k^{\text{F}})^H }_{\mathbf{A}_{\text{FF}}^{W} } \Big] \bm{\xi}_{\text{F}} \\
			& + \left(\bm{\xi}_{\text{B}} \right)^T \Big[ \underbrace{ \sum_{k \in \{ \text{FU},\text{BU} \} } \kappa_k |t_k|^2  \mathbf{R}_k^{\text{B}} \mathbf{W}_{\text{sum}} (\mathbf{R}_k^{\text{B}})^H }_{\mathbf{A}_{\text{BB}}^{W} }  \Big] \bm{\xi}_{\text{B}} \\
			& + 2 \Re \Big\{  \left(\bm{\xi}_{\text{F}} \right)^T \Big[ \underbrace{ \sum_{k \in \{ \text{FU},\text{BU} \} } \kappa_k |t_k|^2 \mathbf{R}_k^{\text{F}} \mathbf{W}_{\text{sum}} ( \mathbf{R}_k^{\text{B}} )^H}_{\mathbf{A}_{\text{FB}}^{W} }   \Big] \bm{\xi}_{\text{B}}  \Big\} \\
			& - 2\sum_{k \in \{ \text{FU},\text{BU} \} }  \Re\Big\{  \left(\bm{\xi}_{\text{F}} \right)^T \!   \kappa_k t_k^* \mathbf{R}_k^{\text{F}} \mathbf{w}_k \!+\! \left(\bm{\xi}_{\text{B}} \right)^T \kappa_k t_k^* \mathbf{R}_k^{\text{B}} \mathbf{w}_k \!  \Big\}, 
		\end{aligned}
	\end{equation}
	and $\mathbf{R}_k^{\chi} = \text{diag}(\mathbf{h}_k^{\chi}) (\mathbf{G}_{\chi}^{\text{PS}} )^T$, $\mathbf{W}_{\text{sum}} = \mathbf{w}_{\text{FU}} \mathbf{w}_{\text{FU}}^H + \mathbf{w}_{\text{BU}} \mathbf{w}_{\text{BU}}^H $. Although the objective function $f_{\bm{\xi}}^{\text{PS}}$ is convex w.r.t. both $\bm{\xi}_{\text{F}}$ and $\bm{\xi}_{\text{B}}$, the unit-norm constraints $\|\bm{\xi}_{\chi}\|^2 = 1$ render the feasible set non-convex. In particular, these constraints restrict the solution to lie on the surface of positive-value unit spheres, which are particularly well-suited for Riemannian manifold optimization techniques. To facilitate this, we define the specific Riemannian manifold $\mathcal{M}_{\xi}$ corresponding to the positive-value unit-modulus constraint as
	\begin{equation}\label{M_chi_set}
		\mathcal{M}_{\xi} = \left\{ \bm{\xi}_{\chi} \in \mathbb{R}^{N \times 1}: \left[ \bm{\xi}_{\chi}  \right]_n \ge 0, \  \left\| \bm{\xi}_{\chi} \right\|^2 = 1 \right\}.
	\end{equation}
	The optimization of $\bm{\xi}_{\chi}$ relies on determining the Riemannian gradient of $f_{\bm{\xi}}^{\text{PS}}$, which is given by
	\begin{equation}\label{gradient_R_PD}
		\nabla_{\mathcal{M}}  f_{\bm{\xi}}^{\text{PS}}(\bm{\xi}_{\chi})  = \nabla  f_{\bm{\xi}}^{\text{PS}} (\bm{\xi}_{\chi})  \!-\!  \text{Tr}\left( (\bm{\xi}_{\chi}) ^T  \nabla  f_{\bm{\xi}}^{\text{PS}} (\bm{\xi}_{\chi})   \right)\! \bm{\xi}_{\chi},
	\end{equation}
	where $\nabla  f_{\bm{\xi}}^{\text{PS}} (\bm{\xi}_{\chi}) $ is the corresponding Euclidean gradient. These expressions can be obtained by
	\begin{subequations}
		\begin{align}
			& \nabla  f_{\bm{\xi}}^{\text{PS}} (\bm{\xi}_{\text{F}}) =  2\Re\Big\{ \mathbf{A}_{\text{FF}}^{W} \bm{\xi}_{\text{F}} +     \mathbf{A}_{\text{FB}}^W \bm{\xi}_{\text{B}} - \!\!\! \sum_{k \in \{ \text{FU},\text{BU} \} } \kappa_k t_k^* \mathbf{R}_k^{\text{F}} \mathbf{w}_k  \Big\},  \\
			& \nabla  f_{\bm{\xi}}^{\text{PS}} (\bm{\xi}_{\text{B}}) =  2\Re\Big\{\mathbf{A}_{\text{BB}}^W  \bm{\xi}_{\text{B}}  +   (\mathbf{A}_{\text{FB}}^W)^T \bm{\xi}_{\text{F}} - \!\!\!  \!\!\!\sum_{k \in \{ \text{FU},\text{BU} \} } \!\!\! \!\!\! \kappa_k t_k^* \mathbf{R}_k^{\text{B}} \mathbf{w}_k    \Big\}.
		\end{align}
	\end{subequations}
	Subsequently, an iterative update of $\bm{\xi}_{\chi}$ is performed by moving along the negative direction of the Riemannian gradient. In order to strictly satisfy both the unit-norm and non-negativity constraints after each update iteration, we employ a projected retraction operation via
	\begin{equation}\label{update_xi_PD}
		\bm{\xi}_{\chi} \gets \frac{\mathcal{P}_+ \big( \bm{\xi}^{\chi} - \varpi_{\xi} \cdot\nabla_{\mathcal{M}}  f_{\bm{\xi}}^{\text{PS}}(\bm{\xi}_{\chi})  \big)}{\big| \mathcal{P}_+ \big( \bm{\xi}^{\chi} - \varpi_{\xi} \cdot\nabla_{\mathcal{M}}  f_{\bm{\xi}}^{\text{PS}}(\bm{\xi}_{\chi})  \big) \big|},
	\end{equation}
	where $\mathcal{P}_{+}(\mathbf{x}) = \max(\mathbf{x}, 0)$ denotes the element-wise projection ensures each element of $\mathbf{x}$ non-negative, and the normalization enforces the unit-norm constraint. The step size $\varpi^{\text{PS}}$ is updated by the Armijo backtracking line search \cite{hosseini2018line}.

	\subsubsection{Subproblem w.r.t. $\mathbf{P}_{\text{PA}}$}
	To facilitate the optimization of PA positions, we adopt the micro-adjustment model where the relative coordinate of the $n$-th PA in direction $\chi$ w.r.t. the nearest input port is given by $x_n^{\chi} = n L_{\text{PA}} + \delta_n^{\chi}$, with $\delta_n^{\chi} \in [-\Delta, \Delta]$ representing the tunable displacement. As established in \cite{gan2025dual}, such micro-positioning, i.e., $\Delta=0.01$ m, mainly changes the phase of the effective channel while causing negligible variations in the amplitude term $(r_{k,n}^{\chi, \text{U}})^{-1}$. Consequently, we can transform the PA position optimization into a tractable phase design problem. By reformulating the channel response to separate the fixed amplitude and tunable phase components, the effective channel vector is expressed as
	\begin{equation}
		\mathbf{h}_{k,\text{eff}} = \sum_{\chi \in \{\text{F}, \text{B} \}} \mathbf{b}_{\chi} \bm{\xi}_{\chi}^T \mathbf{D}(\mathbf{r}_k^{\chi, \text{U}}) \bm{\phi}_k^{\chi},
	\end{equation} 
	where $\mathbf{b}_{\text{F}} =  [\sqrt{\beta_1^{\text{F}}}\! \exp(-j k_g (M\!-\!1) L_{\text{in}}), \! \sqrt{\beta_2^{\text{F}}} \!\exp(-j k_g (M\!-\!2) L_{\text{in}}),\! \dots\!, \sqrt{\beta_M^{\text{F}}} ]^T$, $\mathbf{b}_{\text{B}} = [\sqrt{\beta_1^{\text{B}}} , \sqrt{\beta_2^{\text{B}}}\! \exp(-j k_g  L_{\text{in}}),\dots , \\ \sqrt{\beta_M^{\text{B}}}\! \exp(-j k_g (M\!-\!1) L_{\text{in}}) ]^T$, $\mathbf{D}(\mathbf{r}) = \text{diag}\{ r_1^{-1}, r_2^{-1}, \dots, \\ r_N^{-1} \}$, and the adjustable phase terms are
	\begin{subequations}
		\begin{align}
			& [\bm{\phi}_k^{\text{F}}]_n = \exp(-j k_0 r_{k,n}^{\text{F, U}} - j k_g x_{n}^{\text{F}}), \\
			& [\bm{\phi}_k^{\text{B}}]_n = \exp(-j k_0 r_{k,n}^{\text{B, U}} - j k_g x_{n}^{\text{B}}). 
		\end{align}
	\end{subequations}
	
	Although there exist coupling between $\bm{\phi}_{\text{FU}}^{\text{F}}$ and $\bm{\phi}_{\text{BU}}^{\text{F}}$, and between $\bm{\phi}_{\text{BU}}^{\text{F}}$ and $\bm{\phi}_{\text{BU}}^{\text{B}}$, we can first treat them as decoupled optimization variables and then find the optimal PA positions to approximate the obtained $\bm{\phi}_k^{\chi}$. As such, the optimization problem w.r.t. the phase terms $\bm{\phi}_k^{\chi}$ becomes:
	\begin{subequations}\label{sec4_ES_phi}
		\begin{align}
			& \min_{\bm{\Phi}} \ f_{\bm{\Phi}}^{\text{PS}}(\bm{\Phi}) = \bm{\Phi}^{H}\mathbf{R}_{\Phi} \bm{\Phi}
			-2\Re\!\big\{\mathbf{k}_{\Phi}^{H}\bm{\Phi}\big\},\\
			& \text{s.t.} \quad | [\bm{\phi}_{k}^{\chi}]_n| = 1, \forall n,
		\end{align}
	\end{subequations}
	where $\bm{\Phi}=
	\Big[
	(\bm{\phi}_{\text{FU}}^{\text{F}})^{T},
	(\bm{\phi}_{\text{FU}}^{\text{B}})^{T},
	(\bm{\phi}_{\text{BU}}^{\text{F}})^{T},
	(\bm{\phi}_{\text{BU}}^{\text{B}})^{T}
	\Big]^{T}\in\mathbb{C}^{4N \times 1}$, $\mathbf{R}_{\Phi}=\text{blkdiag}\big(\mathbf{R}_{\text{FU}},\mathbf{R}_{\text{BU}}\big)$, $\mathbf{R}_{k}=\kappa_{k}|t_{k}|^{2}\eta \mathbf{K}_{k}^{H}\mathbf{W}_{\text{sum}}\mathbf{K}_{k}$, $\mathbf{K}_k =  [ \mathbf{b}_{\text{F}} \bm{\xi}_{\text{F}}^T \mathbf{D}(\mathbf{r}_k^{\text{F}, \text{U}}), \mathbf{b}_{\text{B}} \bm{\xi}_{\text{B}}^T \mathbf{D}(\mathbf{r}_k^{\text{B}, \text{U}})  ] \in \mathbb{C}^{M \times 2N} $, $\mathbf{k}_{\Phi}= [\kappa_{\text{FU}}t_{\text{FU}}\mathbf{w}_{\text{FU}}^H \mathbf{K}_{\text{FU}},  \kappa_{\text{BU}}t_{\text{BU}} \mathbf{w}_{\text{BU}}^H  \mathbf{K}_{\text{BU}} ]^T$. In a similar way, we employ the Riemannian gradient descent to optimize the phase vector $\bm{\Phi}$. In particular, the Euclidean gradient of the objective function is given by $\nabla f_{\bm{\Phi}}^{\text{PS}}(\bm{\Phi}) = 2 \mathbf{R}_{\Phi} \bm{\Phi} - 2 \mathbf{k}_{\Phi}$. Based on this, the Riemannian gradient is calculated by
	\begin{equation}\label{rie_gra_PS}
		\nabla_{\mathcal{M}} f_{\bm{\Phi}}^{\text{PS}}(\bm{\Phi}) = \nabla f_{\bm{\Phi}}^{\text{PS}}(\bm{\Phi}) - \text{Tr} \left\{ (\bm{\Phi})^T   \nabla f_{\bm{\Phi}}^{\text{PS}}(\bm{\Phi})  \right\} \bm{\Phi},
	\end{equation}
	The iterative update is then performed via a retraction mapping that projects onto the unit module:
	\begin{equation}\label{update_phi_PD}
		[\bm{\Phi}]_n \gets \frac{[\bm{\Phi}]_n - \varpi_{\Phi} \cdot  \nabla_{\mathcal{M}} f_{\bm{\Phi}}^{\text{PS}}(\bm{\Phi})  }{\big| [\bm{\Phi}]_n - \varpi_{\Phi} \cdot  \nabla_{\mathcal{M}} f_{\bm{\Phi}}^{\text{PS}}(\bm{\Phi})  \big|},
	\end{equation}
	where the step size $\varpi_{\Phi}$ is also determined by the Armijo backtracking line search rule. Based on the obtained converged variables $\bm{\Phi}$, the optimization of PA position is solved by the least-squares phase alignment problem as
	\begin{equation}\label{update_d_PS}
		d_n^{\chi} = \arg \min_{d \in [-\Delta, \Delta]} \sum_{k \in \{\text{FU}, \text{BU}\}} \left| [\bm{\phi}_{k}^{\chi}]_n - e^{-j (k_0 r_{k,n}^{\chi}(d) + k_g d)} \right|^2.
	\end{equation}
	This one-dimensional minimization problem is efficiently solved via a grid search over the region $[-\Delta, \Delta]$. Then, based on the optimal $d_n^{\chi}$ for $\forall n,\chi$, the position coordinates of PAs are given by $\mathbf{p}_n^{\text{FPA}} = [\frac{M-1}{2} L_{\text{IN}} + n L_{\text{PA}} + d_n^{\text{F}} , 0]$ and $\mathbf{p}_n^{\text{BPA}} = [\frac{1-M}{2} L_{\text{IN}} - n L_{\text{PA}} + d_n^{\text{B}}, 0]$.

	\subsubsection{Subproblem w.r.t. $\mathbf{\beta}_{\chi}$}\label{subsec:beta}
	
	Fixing $\{\mathbf{W},\bm{\Sigma}_{\chi},\mathbf{P}_{\text{PA}}\}$ and dropping all terms independent of $\bm{\beta}_{\chi}$ from \eqref{sec4_problem_PS_all}, the subproblem w.r.t. the power splitting ratio vectors $\bm{\beta}_{\chi} = [\sqrt{\beta_1^{\chi}},\sqrt{\beta_2^{\chi}}, \dots,\sqrt{\beta_M^{\chi}}]^T$ for $\chi \in \{ \text{F}, \text{B} \}$ is reduced to
	\begin{subequations}\label{sec4_PD_beta}
		\begin{align}
			\min_{\bm{\beta}_{\text{F}},\,\bm{\beta}_{\text{B}}}\quad
			& f_{\bm{\beta}}^{\text{PS}} \label{sec4_PD_beta_obj}\\
			\text{s.t.}\quad
			& [\bm{\beta}_{\text{F}}]_m^2 + [\bm{\beta}_{\text{B}}]_m^2 = 1,\ \forall m, \label{sec4_PD_beta_cpl}\\
			& [\bm{\beta}_{\chi}]_m \in [0,1],\ \forall m, \label{sec4_PD_beta_ps}
		\end{align}
	\end{subequations}
	where the objective function is given by
	\begin{equation}\label{eq:f_beta_def}
		\begin{aligned}
			f_{\bm{\beta}}^{\text{PS}}
			=
			\sum_{k\in\{\text{FU},\text{BU}\}} &
			\kappa_k|t_k|^2 (\alpha_k^{\text{F}} \bm{\beta}_{\text{F}}^T \mathbf{B}_{\text{F}} + \alpha_k^{\text{B}}\bm{\beta}_{\text{B}}^T \mathbf{B}_{\text{B}} ) \mathbf{W}_{\text{sum}} \\
			& \times ( \underbrace{ (\alpha_k^{\text{F}})^* \mathbf{B}_{\text{F}}^* \bm{\beta}_{\text{F}} + (\alpha_k^{\text{B}})^* \mathbf{B}_{\text{B}}^* \bm{\beta}_{\text{B}} }_{\mathbf{h}_{k,\text{eff}}^*} ) \\
			& -
			2\Re\!\big\{\kappa_k t_k^{*} (\alpha_k^{\text{F}} \bm{\beta}_{\text{F}}^T \mathbf{B}_{\text{F}} + \alpha_k^{\text{B}}\bm{\beta}_{\text{B}}^T \mathbf{B}_{\text{B}} ) \mathbf{w}_k\big\}
			\Big),
		\end{aligned}
	\end{equation}
	$\alpha_k^{\chi} = \bm{\xi}_{\chi} \text{diag}(\mathbf{h}_k^{\chi})  \mathbf{b}_{\chi}$ and $\mathbf{B}_{\chi} = \text{diag}\{ \mathbf{b}_{\chi} \}$. To tackle the coupled quadratic constraints \eqref{sec4_PD_beta_cpl},  we employ a polar coordinate transformation, i.e.,
	\begin{equation}\label{eq:theta_param}
		[\bm{\beta}_{\text{F}}]_m=\cos(\theta_m),\quad
		[\bm{\beta}_{\text{B}}]_m=\sin(\theta_m).
	\end{equation}
	Accordingly, the constraint~\eqref{sec4_PD_beta_ps} can be transformed into the constraints of $\theta_m \in [0, \pi/2], \ \forall m$ for the PS. Consequently, the optimizations of $\bm{\beta}_{\text{F}}$ and $\bm{\beta}_{\text{B}}$ are transformed into a bound-constrained smooth minimization problem w.r.t. $\bm{\theta} = [\theta_1, \theta_2,\dots,\theta_M]^T$. We solve this efficiently using a quasi-Newton method~\cite{nocedal2006numerical} equipped with the derived analytical gradient as
	\begin{equation}\label{eq:grad_theta}
		\nabla_{\bm{\theta}} f_{\bm{\beta}}^{\text{PS}}
		=
		\big(\nabla_{\bm{\beta}_{\text{B}}} f_{\bm{\beta}}^{\text{PS}}\big)\odot \cos(\bm{\theta})
		-
		\big(\nabla_{\bm{\beta}_{\text{F}}} f_{\bm{\beta}}^{\text{PS}}\big)\odot \sin(\bm{\theta}),
	\end{equation}
	where the Euclidean gradients are given by
	\begin{subequations}\label{eq:grad_beta_FB}
		\begin{align}
			&\nabla_{\bm{\beta}_{\text{F}}} f_{\bm{\beta}}^{\text{PS}}
			\!=
			\!\!\! \!\!\!  \sum_{k \in \{ \text{FU},\text{BU} \}} \!\!\! \!\!\!  2 \Re\!\big\{\alpha_{k,\text{F}}\mathbf{B}_{\text{F}}^{T} (\kappa_k|t_k|^2\,\mathbf{W}_{\text{sum}} \mathbf{h}_{k,\text{eff}}^{*}\!-\! \kappa_k t_k^{*}\mathbf{w}_k)  \big\},\\
			& \nabla_{\bm{\beta}_{\text{B}}} f_{\bm{\beta}}^{\text{PS}}
			\!= \!\!\! \!\!\! 
			\sum_{k \in \{ \text{FU},\text{BU} \}} \!\!\! \!\!\! 2 \Re\!\big\{\alpha_{k,\text{B}}\mathbf{B}_{\text{B}}^{T}(\kappa_k|t_k|^2\,\mathbf{W}_{\text{sum}} \mathbf{h}_{k,\text{eff}}^{*}\!-\! \kappa_k t_k^{*}\mathbf{w}_k) \big\}.
		\end{align}
	\end{subequations}
	Based on the derived first-order gradient in \eqref{eq:grad_theta}, we employ the sequential quadratic programming (SQP) approach to derive the approximated Hessian matrix with low complexity. By performing an efficient quasi-Newton method to update the variables $\bm{\theta}$ until convergence. Then, the power splitting ratios can be derived through \eqref{eq:theta_param}. 
	
	\begin{algorithm}[t]
		\caption{Proposed Alternating Optimization Algorithm to Solve Problem~\eqref{sec3_PSDS_problem} for PS and DS}
		\label{algorithm_PS_DS}
		\begin{algorithmic}[1]
			\STATE Initialize the optimization variables.
			\REPEAT
			\STATE Update $\mathbf{t}$ and $\bm{\kappa}$ as~\eqref{update_t_kappa_PS}.
			\STATE Update $\mathbf{W}$ as~\eqref{update_W_PS} with $\lambda^{\text{opt}}$.
			\STATE Update $\bm{\Sigma}_{\chi}$ by solving problem~\eqref{sec4_ES_xi} through Riemannian optimization method.
			\STATE Update $\mathbf{P}_{\text{PA}}$ by solving problems~\eqref{sec4_ES_phi} and \eqref{update_d_PS}.
			\vspace{1mm}
			\STATE 
			\emph{Protocol-specific update of $\bm{\xi}_{\chi}$}\\
			\textbf{For PS:} Update $\bm{\xi}_{\chi}$ by solving problem~\eqref{sec4_PD_beta} through quasi-Newton method.\\
			\textbf{For DS:} Update $\bm{\xi}_{\chi}$ by solving the penalty-based problem~\eqref{eq:obj_pen} with $\rho$ through quasi-Newton method.
			\vspace{1mm}
			\UNTIL{Convergence or reaching the maximum number of iteration $I_{\max}$.}
		\end{algorithmic}
	\end{algorithm}
	
	\subsection{The Proposed Solution for DS}
	Compared with the PS protocol, the DS protocol only replaces the continuous power splitting ratio design with a binary constraint. This renders the problem~\eqref{sec4_PD_beta} involving mixed-integer constraints, which is generally NP-hard and intractable to solve directly. To address this challenge, we extend the optimization framework developed for the PS protocol by incorporating a penalty-based term. This approach can efficiently relax the binary constraints into continuous ones. Specifically, for any continuous variable $x \in [0, 1]$, the inequality $x - x^2 \ge 0$ holds, where the equality is achieved if and only if $x \in \{0, 1\}$. Leveraging this property, we construct a penalized objective function by augmenting the original objective function~\eqref{sec4_PD_beta_obj} with a penalty term:
	\begin{subequations} \label{eq:obj_pen}
		\begin{align}
			\min_{\bm{\beta}_{\text{F}},\,\bm{\beta}_{\text{B}}}\
			& \mathcal{L}(\bm{\beta}, \rho) = f_{\bm{\beta}}^{\text{DS}} + \rho \!\! \sum_{m=1}^{M} \sum_{\chi \in \{\text{F}, \text{B}\}} \!\! \left( [\bm{\beta}_{\chi}]_m\!-\! [\bm{\beta}_{\chi}]_m^2 \right) \\
			\text{s.t.}\quad
			& [\bm{\beta}_{\text{F}}]_m^2 + [\bm{\beta}_{\text{B}}]_m^2 = 1,\ \forall m,
		\end{align}
	\end{subequations}
	where $\rho > 0$ is a penalty parameter, and $f_{\bm{\beta}}^{\text{DS}}$ is the same as in \eqref{eq:f_beta_def}. Consequently, the optimization of the DS protocol is performed by iteratively minimizing $\mathcal{L}(\bm{\theta}, \rho)$ while gradually increasing the value of $\rho$ in each outer iteration. As $\rho \to \infty$, the penalty term dominates, enforcing the solution to satisfy the binary constraints $[\bm{\beta}_{\chi}]_m \in \{0,1\}$. Thus, the value of $\rho$ can increase with the number of iteration, such as $\rho^{\text{iter+1}} = c_{\rho} \cdot \rho^{\text{iter}}$ with $c_{\rho} > 1$. Since the penalty-based objective function~\eqref{eq:obj_pen} remains differentiable w.r.t. $\bm{\theta}$, we can continue to apply the quasi-Newton method~\cite{nocedal2006numerical}. In particular, the gradient can be written as
	\begin{equation} \label{eq:grad_theta_DS}
		\nabla_{\bm{\theta}} \mathcal{L} = \nabla_{\bm{\theta}} f_{\bm{\beta}}^{\text{DS}} + \rho \cdot \nabla_{\bm{\theta}} P(\bm{\theta}),
	\end{equation}
	where $P(\bm{\theta}) = \sum_{m=1}^M  (\cos \theta_m - \cos^2 \theta_m) + (\sin \theta_m - \sin^2 \theta_m)$, $\nabla_{\bm{\theta}} f_{\bm{\beta}}^{\text{DS}}$ is identical to the PS gradient given in \eqref{eq:grad_beta_FB}. By substituting this into the SQP solver, the algorithm efficiently identifies a high-quality binary solution.
	
	\begin{remark}
		The proposed alternating optimization algorithm to maximize the sum rate of C-PASS for PS and DS is summarized in \emph{Algorithm}~\ref{algorithm_PS_DS}. In particular, the iterative updates of the auxiliary variables $\mathbf{t}$ and $\bm{\kappa}$ establish the equivalence between the original sum-rate maximization problem and the tractable WMMSE formulation. It is therefore beneficial to update these auxiliary variables immediately following the optimization of each variable. Moreover, it can be noticed that PS and DS share the same optimization structure, differing only in the update of the power splitting ratios $\bm{\xi}_{\chi}$ as continuous or discrete variables.
	\end{remark}

	\subsection{The Proposed Solution for TS}
	For the TS protocol, user transmissions are orthogonalized in the time domain. Consequently, the optimal transmit precoding strategy is the maximum ratio transmission (MRT)\cite{9617121}, which is given by
	\begin{align}
		\mathbf{w}_{\text{FU}} = \sqrt{P_T} \frac{ \mathbf{h}_{\text{FU, eff}}^* }{\left\| \mathbf{h}_{\text{FU, eff}} \right\|},\quad \mathbf{w}_{\text{BU}} = \sqrt{P_T} \frac{ \mathbf{h}_{\text{BU, eff}}^* }{\left\| \mathbf{h}_{\text{BU, eff}} \right\|}
	\end{align}
	where the effective channel contains only one-way path for either FP or BP link as $\mathbf{h}_{\text{FU,eff}} = \mathbf{G}_{\text{F}}^{\text{TS}} \bm{\Sigma}_{\text{F}} \mathbf{h}_{\text{FU}}^{\text{F}}$ and $\mathbf{h}_{\text{BU,eff}} = \mathbf{G}_{\text{B}}^{\text{TS}} \bm{\Sigma}_{\text{B}} \mathbf{h}_{\text{BU}}^{\text{B}}$. Substituting these closed-form precoders into the sum-rate expression for $k$-user \eqref{sec3_R_TS}, the optimization objective simplifies to a function of the effective channel gain as
	\begin{equation} \label{eq:R_TS_simplified}
		R_k^{\text{TS}} = \mu_k \log_2 \left( 1 + \frac{P_T}{\mu_k N_0} \left\| \mathbf{h}_{k,\text{eff}} \right\|^2 \right),
	\end{equation}
	Although the optimization of the precoding vectors is resolved in closed form, maximizing sum-rate \eqref{eq:R_TS_simplified} w.r.t. the remaining coupled variables $\{ \bm{\mu}, \bm{\Sigma}_{\chi}, \mathbf{P}_{\text{PA}} \}$ remains a non-convex problem. To address this tractably and maintain consistency with the optimization framework proposed for the PS/DS protocols, we transform the objective function \eqref{eq:R_TS_simplified} into an equivalent tractable form in the following lemma.
	
	\begin{lemma}
		The sum-rate maximization function~\eqref{sec3_TS_problem_function} for TS is equivalent to the weighted error minimization function as
		\begin{align}\label{sec4_problem_TS}
			\min_{\bm{\hat{\kappa}},\mathbf{\hat{T}},\bm{\mu},\bm{\Xi}_{\chi}, \mathbf{P}_{\text{PA}}} \
			\sum_{k \in \{ \text{FU},\text{BU} \} }  \frac{\mu_k \hat{\kappa}_k \hat{\epsilon}_k}{\ln 2} - \mu_k \log_2(\hat{\kappa}_k),
		\end{align}
		where $\hat{\bm{\epsilon}} = \left[ \hat{\epsilon}_{\text{FU}},\hat{\epsilon}_{\text{BU}} \right]$, $\hat{\bm{\kappa}} = \left[ \hat{\kappa}_{\text{FU}}, \hat{\kappa}_{\text{BU}} \right]$ and $\mathbf{\hat{T}} = \left[ \mathbf{\hat{t}}_{\text{FU}}, \mathbf{\hat{t}}_{\text{BU}} \right]$ are the auxiliary vector and matrix, and the equivalent error expression is given by
		\begin{equation}
			\hat{\epsilon}_k = \left( 1- \hat{\mathbf{t}}_k^H \mathbf{h}_{k,\text{eff}} \right)  \left( 1- \mathbf{h}_{k,\text{eff}}^H  \hat{\mathbf{t}}_k\right) + \frac{\mu_k N_0}{P_T} \hat{\mathbf{t}}_k^H \hat{\mathbf{t}}_k.
		\end{equation}
		\begin{proof}
			Please refer to the proof derivation of [\emph{Theorem 1},~\citen{shi2011iteratively}].
		\end{proof}
	\end{lemma}
	Since the  objective function~\eqref{sec4_problem_TS} is strictly convex w.r.t. $\bm{\hat{\kappa}}$ and $\mathbf{\hat{T}}$ with no constraints. Thus, the optimal $\bm{\hat{\kappa}}$ and $\mathbf{\hat{T}}$ can be derived by setting their first-order derivatives to zero, i.e.,
	\begin{subequations}\label{update_t_kappa_TS}
		\begin{align}
			& \hat{\mathbf{t}}_k^{\text{opt}} = \Big( \mathbf{h}_{k,\text{eff}} \mathbf{h}_{k,\text{eff}}^H + \frac{\mu_k N_0}{P_T} \mathbf{I}_M \Big)^{-1}  \mathbf{h}_{k,\text{eff}}, \\
			& \hat{\kappa}_k^{\text{opt}} = \hat{\epsilon}_k^{-1}.
		\end{align}
	\end{subequations}
	Substituting the optimal $\bm{\hat{\kappa}}$ and $\mathbf{\hat{T}}$ into \eqref{sec4_problem_TS}, the optimization problem for TS is reduced to 
	\begin{subequations}\label{sec4_problem_TS2}
		\begin{align}
			\min_{\bm{\mu},\bm{\Xi}_{\chi}, \mathbf{P}_{\text{PA}}}
			&  \ f_{\text{all}}^{\text{TS}}( \bm{\mu},\bm{\Xi}_{\chi}, \mathbf{P}_{\text{PA}}) , \label{sec4_problem_TS2_function}\\
			\text{s.t.}\quad
			& \mu_{\text{FU}}+\mu_{\text{BU}}=1, \\
			& \mu_{\text{FU}},\mu_{\text{BU}} \in [0,1] ,\\
			& \|\mathbf{W}\|^2 \le P_T, \\
			& \|\bm{\Sigma}_{\chi}\|^2 = 1, \forall \chi, \\
			& \mathbf{P}_{\text{PA}} \in \mathcal{G}(\mathbf{P}_{\text{PA}}),
		\end{align}
	\end{subequations}
	where the objective function~\eqref{sec4_problem_TS2_function} can be rewritten as
	\begin{equation}\label{sec4_TS_obj_all}
		\begin{aligned}
			f_{\text{all}}^{\text{TS}}( \bm{\mu},\bm{\Xi}_{\chi},\! \mathbf{P}_{\text{PA}}) =\!\!\!\!\sum_{k \in \{ \text{FU},\text{BU} \} }\!\!& \frac{\mu_k \hat{\kappa}_k}{\ln 2} \! \left( \!1 \!-\! \hat{\mathbf{t}}_k^H \mathbf{h}_{k,\text{eff}} \right) \! \left( \!1\!-\! \mathbf{h}_{k,\text{eff}}^H  \hat{\mathbf{t}}_k \right) 	\\
			& + \frac{\mu_k^2 \hat{\kappa}_k N_0}{P_T \ln 2} \hat{\mathbf{t}}_k^H \hat{\mathbf{t}}_k \!-\!  \mu_k \log_2(\hat{\kappa}_k).
		\end{aligned}
	\end{equation}
	
	\begin{remark}
		Under the TS protocol, C-PASS serves the communication users sequentially rather than simultaneously. Specifically, during the $\mu_{\text{FU}}$ period, the C-PASS activates only the FP link to serve the FU, while the reverse applies during the $\mu_{\text{BU}}$ period. Consequently, the sum-rate maximization problem can be decoupled into two independent single-user rate maximization subproblems for TS. This implies that the optimization problem of $\bm{\Xi}_{\chi}$ and $\mathbf{P}_{\text{PA}}$ for TS is a special case of the PS and DS protocols. Therefore, to avoid redundancy, we omit the repetitive derivations and just outline the necessary adaptations for the TS implementation. The overall algorithm for optimizing C-PASS under the TS protocol is summarized in \emph{Algorithm}~\ref{algorithm_TS}.
	\end{remark}
	
	\begin{algorithm}[t]
		\caption{Proposed Alternating Optimization Algorithm to Solve Problem~\eqref{sec3_TS_problem} for TS}
		\label{algorithm_TS}
		\begin{algorithmic}[1]
			\STATE Initialize the optimization variables.
			\REPEAT
			\STATE Update $\mathbf{\hat{T}}$ and $\bm{\hat{\kappa}}$ as~\eqref{update_t_kappa_TS}.
			\STATE Update $\bm{\mu}$ as \eqref{update_mu_TS}.
			\STATE Update $\bm{\Sigma}_{\chi}$ by solving problem~\eqref{xi_TS_pro} through Riemannian optimization method.
			\STATE Update $\mathbf{P}_{\text{PA}}$ by solving problems~\eqref{eq:TS_phi_quad} and \eqref{update_d_TS}.
			\UNTIL{Convergence or reaching the maximum number of iteration $I_{\max}$.}
		\end{algorithmic}
	\end{algorithm}

	\subsubsection{Subproblem w.r.t. $\mathbf{\mu}_k$}
	Under the TS protocol, the time allocation ratios satisfy $ \mu_{\text{FU}}+ \mu_{\text{BU}}=1$.  By substituting $\mu_{\text{BU}}=1-\mu_{\text{FU}}$ into the objective function~\eqref{sec4_TS_obj_all} and discarding all constants independent of $\mu_{\text{FU}}$, the optimization of $\mu_{\text{FU}}$ reduces to a one-dimensional quadratic minimization over the closed interval $[0,1]$, i.e.,
	\begin{subequations}
		\begin{align}
			\min_{\mu_{\text{FU}}} & \quad A_{\mu}^{\text{FU}} \left( \mu_{\text{FU}} \right)^2 + B_{\mu}^{\text{FU}} \mu_{\text{FU}} + C_{\mu}^{\text{FU}}  \\ 
			\text{s.t.} &  \quad  \mu_{\text{FU}} \in [0,1],\label{TS_mu_con}
		\end{align}
	\end{subequations}
	where the coefficients are given by
	\begin{align}
		&A_{\mu}^{\text{FU}} \! = \! \sum_{k \in \{ \text{FU},\text{BU} \} }  \frac{ \hat{\kappa}_k N_0}{P_T} \| \hat{\mathbf{t}}_k  \|^2,  \\
		&B_{\mu}^{\text{FU}} \!  =\!   \hat{\kappa}_{\text{FU}} \!\left|  \hat{\mathbf{t}}_{\text{FU}}^H \! ( \mathbf{G}_{\text{F}}^{\text{TS}} )^* \bm{\Xi}_{\text{F}}  	( \mathbf{h}_{\text{FU}}^{\text{F}} )^* \!  \right|^2\! \!  -\!   \hat{\kappa}_{\text{BU}} \!\left| \hat{\mathbf{t}}_{\text{BU}}^H \! (\mathbf{G}_{\text{B}}^{\text{TS}} )^*\bm{\Xi}_{\text{B}} \! ( \mathbf{h}_{\text{BU}}^{\text{B}} )^*\!   \right|^2  \notag\\
		&\!-\! 2\hat{\kappa}_{\text{FU}}\Re\! \left\{\! \hat{\mathbf{t}}_{\text{FU}}^H \! ( \mathbf{G}_{\text{F}}^{\text{TS}} )^* \bm{\Xi}_{\text{F}} \! 	( \mathbf{h}_{\text{FU}}^{\text{F}} )^* \! \right\} \!\! + \! 2\hat{\kappa}_{\text{BU}}\Re\! \left\{\! \hat{\mathbf{t}}_{\text{BU}}^H \! (\! \mathbf{G}_{\text{B}}^{\text{TS}} )^*\bm{\Xi}_{\text{B}}  	( \mathbf{h}_{\text{BU}}^{\text{B}} )^* \! \right\}  \notag \\
		& + 2 \frac{ \hat{\kappa}_{\text{FU}} N_0}{P_T} \| \hat{\mathbf{t}}_{\text{FU}} \|^2 - 2 \frac{ \hat{\kappa}_{\text{BU}} N_0}{P_T} \| \hat{\mathbf{t}}_{\text{BU}}\|^2, \\
		& C_{\mu}^{\text{FU}} = \hat{\kappa}_{\text{BU}} \!\left| \hat{\mathbf{t}}_{\text{BU}}^H  ( \mathbf{G}_{\text{B}}^{\text{TS}} )^* \bm{\Xi}_{\text{B}} \! 	( \mathbf{h}_{\text{BU}}^{\text{B}} )^*  \right|^2 + \frac{ \hat{\kappa}_{\text{BU}} N_0}{P_T} \| \hat{\mathbf{t}}_{\text{BU}}  \|^2 \notag \\ 
		& \quad - 2 \hat{\kappa}_{\text{BU}}\Re \left\{ \hat{\mathbf{t}}_{\text{BU}}^H \! ( \mathbf{G}_{\text{B}}^{\text{TS}})^* \bm{\Xi}^{\text{B}} \! 	( \mathbf{h}_{\text{BU}}^{\text{B}} )^*  \right\}.
	\end{align}
	Since the weights $\hat{\kappa}_k$, noise power $N_0$, and transmit power $P_T$ are all positive, the quadratic coefficient satisfies $A_{\mu}^{\text{FU}} > 0$. This indicates that the unconstrained global minimum occurs at the point $- B_{\mu}^{\text{F}} / (2 A_{\mu}^{\text{F}} )$ . Taking the constraints~\eqref{TS_mu_con} into account, the optimal time allocation for the FU is derived by 
	\begin{equation}\label{update_mu_TS}
		\mu_{\text{FU}}^{\text{opt}} = \left\{\begin{matrix}
			0,  &\text{if} \ -\frac{B_{\mu}^{\text{F}}}{2A_{\mu}^{\text{F}}}<0 \\
			-\frac{B_{\mu}^{\text{F}}}{2A_{\mu}^{\text{F}}}, & \text{if} \ -\frac{B_{\mu}^{\text{F}}}{2A_{\mu}^{\text{F}}} \in [0,1]  \\
			1, & \text{if} \ -\frac{B_{\mu}^{\text{F}}}{2A_{\mu}^{\text{F}}}>1.
		\end{matrix}\right.
	\end{equation}
	Accordingly, the optimal time allocation for the BU is determined by $\mu_{\text{BU}}^{\text{opt}} = 1 - \mu_{\text{FU}}^{\text{opt}}$.

	\subsubsection{Subproblem w.r.t. $\mathbf{\Xi}^{\text{C}}$}
	The optimization w.r.t. $\bm{\Xi}^{\text{C}}$ can be decoupled into two independent subproblems w.r.t. $\bm{\xi}^{\text{F}}$ and $\bm{\xi}^{\text{B}}$, which can be written as
	\begin{subequations}\label{xi_TS_pro}
		\begin{align}
			\min_{\bm{\xi}_{\chi}} & \quad f_{\bm{\xi}_{\chi}}^{\text{TS}}   \\
			\text{s.t.} & \quad  \left\|  \bm{\xi}_{\chi} \right\|^2 = 1,
		\end{align}
	\end{subequations}
	where the objective function for $\chi \in \{ \text{F}, \text{B}\}$ are given by
	\begin{subequations}
		\begin{align}
			f_{\bm{\xi}_{\text{F}}}^{\text{TS}}  = &  \bm{\xi}_{\text{F}}^T \text{diag}(\mathbf{h}_{\text{FU}}^{\text{F}}) (\mathbf{G}_{\text{F}}^{\text{TS}})^T \hat{\mathbf{t}}_{\text{FU}}^* \hat{\mathbf{t}}_{\text{FU}}^T (\mathbf{G}_{\text{F}}^{\text{TS}})^* \text{diag}((\mathbf{h}_{\text{FU}}^{\text{F}})^*) \bm{\xi}_{\text{F}} \notag\\
			& - 2 \Re \{ \bm{\xi}_{\text{F}}^T \text{diag}(\mathbf{h}_{\text{FU}}^{\text{F}}) (\mathbf{G}_{\text{F}}^{\text{TS}})^T \hat{\mathbf{t}}_{\text{FU}}^*  \} \\
			f_{\bm{\xi}_{\text{B}}}^{\text{TS}}  = &  \bm{\xi}_{\text{B}}^T \text{diag}(\mathbf{h}_{\text{BU}}^{\text{B}}) (\mathbf{G}_{\text{B}}^{\text{TS}})^T \hat{\mathbf{t}}_{\text{BU}}^* \hat{\mathbf{t}}_{\text{BU}}^T (\mathbf{G}_{\text{B}}^{\text{TS}})^* \text{diag}((\mathbf{h}_{\text{BU}}^{\text{B}})^*) \bm{\xi}_{\text{B}} \notag \\
			& - 2 \Re \{ \bm{\xi}_{\text{B}}^T \text{diag}(\mathbf{h}_{\text{BU}}^{\text{B}}) (\mathbf{G}_{\text{B}}^{\text{TS}})^T \hat{\mathbf{t}}_{\text{BU}}^*  \}
		\end{align}
	\end{subequations}
	It is worth noting that the constraints on $\bm{\xi}_{\chi}$ are on the same manifold $\mathcal{M}_{\xi}$ in \eqref{M_chi_set}. In each iteration, we first compute the Euclidean gradient of $f_{\bm{\xi}_{\chi}}^{\text{TS}}$ as
	\begin{equation}
		\begin{aligned}
			\nabla f_{\bm{\xi}_{\chi}}^{\text{TS}} = & 2 \Re \left\{ \text{diag}(\mathbf{h}_{k}^{\chi}) (\mathbf{G}_{\chi}^{\text{TS}})^T \hat{\mathbf{t}}_{k}^* \hat{\mathbf{t}}_{k}^T (\mathbf{G}_{\chi}^{\text{TS}})^* \text{diag}((\mathbf{h}_{k}^{\chi})^*) \right\} \bm{\xi}_{\chi} \\
			& - 2 \Re \left\{ \text{diag}(\mathbf{h}_{k}^{\chi}) (\mathbf{G}_{\chi}^{\text{TS}})^T \hat{\mathbf{t}}_{k}^* \right\},
		\end{aligned}
	\end{equation}
	and then obtain the Riemannian gradient as in \eqref{gradient_R_PD}. The variable $\bm{\xi}_{\chi}$ is subsequently updated along the negative Riemannian gradient with projected retraction, as in \eqref{update_xi_PD}.

	\subsubsection{Subproblem w.r.t. $\mathbf{P}_{\text{PA}}$}
	Under the TS protocol, the optimization of PA positions only involves the adjustment of 
	$\bm{\phi}^{\text{F}}$ and $\bm{\phi}^{\text{B}}$. In this case, the optimization problem of $\bm{\Phi} = [(\bm{\phi}^{\text{F}})^{T},(\bm{\phi}^{\text{B}})^{T}]^{T}$ can be written as
	\begin{subequations}\label{eq:TS_phi_quad}
		\begin{align}
			\min_{\bm{\Phi}}
			&	\ f_{\bm{\Phi}}^{\text{TS}}(\bm{\Phi})
			= 
			\bm{\Phi}^{H}\mathbf{\hat{R}}_{\Phi} \bm{\Phi}-2\Re\!\big\{\mathbf{\hat{k}}_{\Phi}^{H}\bm{\Phi}\big\}\\
			\text{s.t.} & \quad |[\bm{\Phi}]_n|=1,\ \forall n,
		\end{align}
	\end{subequations}
	where $\mathbf{\hat{R}}_{\Phi} =  \text{blkdiag}\big\{ \mathbf{\hat{R}}_{\text{FU}}, \mathbf{\hat{R}}_{\text{BU}} \big\}$, $\mathbf{\hat{R}}_k = (\mathbf{\hat{K}}_k)^{H} \hat{\mathbf{t}}_k \hat{\mathbf{t}}_k^H   \mathbf{\hat{K}}_k$, $\mathbf{\hat{K}}_{\text{FU}} = \eta \mathbf{\hat{b}}_{\text{F}} \bm{\xi}_{\text{F}}^T \mathbf{D}(\mathbf{r}_{\text{FU}}^{\text{F,U}})$, $\mathbf{\hat{K}}_{\text{BU}} =\eta \mathbf{\hat{b}}_{\text{B}} \bm{\xi}_{\text{B}}^T \mathbf{D}(\mathbf{r}_{\text{BU}}^{\text{B,U}}) $, $\mathbf{b}_{\text{F}}^{\text{TS}} =  [ \exp(-j k_g (M\!-\!1) L_{\text{in}}), \exp(-j k_g (M\!-\!2) L_{\text{in}}),\! \dots\!,1]^T$, $\mathbf{b}_{\text{B}}^{\text{TS}} = [1 ,  \exp(-j k_g  L_{\text{in}}),\dots ,  \exp(-j k_g (M\!-\!1) L_{\text{in}}) ]^T$, and $\mathbf{\hat{k}}_{\Phi} = [\mathbf{\hat{t}}_{\text{FU}}^T \mathbf{\hat{K}}_{\text{FU}}^*, \mathbf{\hat{t}}_{\text{BU}}^T \mathbf{\hat{K}}_{\text{BU}}^* ]^T$. Then, the Euclidean gradient of \eqref{eq:TS_phi_quad} is given by
	\begin{equation}\label{eq:TS_phi_gradE}
		\nabla f_{\bm{\Phi}}^{\text{TS}}(\bm{\Phi})=2\mathbf{\hat{R}}_{\Phi} \bm{\Phi}-2\mathbf{\hat{k}}_{\Phi},
	\end{equation}
	which yields the Riemannian gradient as in \eqref{rie_gra_PS}. After updating $\bm{\Phi}$ with \eqref{update_phi_PD} until convergence, the position displacement of PA $d_n^{\chi}$ can be determined as
	\begin{equation}\label{update_d_TS}
		d_n^{\chi} = \arg \min_{d \in [-\Delta, \Delta]} \left| [\bm{\phi}_{k}^{\chi}]_n - e^{-j (k_0 r_{k,n}^{\chi}(d) + k_g d)} \right|^2,
	\end{equation}
	This reduced single-user objective is efficiently solved via the grid search procedure described previously, and then yielding the PA positions.

	\section{Numerical Results}\label{sec:simulation}
	In this section, numerical results are provided to illustrate the advantages of the C-PASS aided wireless communications and validate the effectiveness of the proposed algorithms under the PS, DS, and TS protocols.
	
	\subsection{Simulation Setup}
	We consider a mmWave scenario with the carrier frequency of $f_c = 28$ GHz. Then, the wavelength, wavenumber, and effective channel gain in the free space can be calculated by $\lambda_0 = c/f_c$, $k_0 = 2\pi/ \lambda_0$, and $\eta = \lambda_0/(4\pi)$, with $c = 3 \times 10^8$ m/s denoting the speed of light. The effective refractive index $n_{\text{eff}} = 1.4$. Unless stated otherwise, the simulation setup is given in Table~\ref{tab:simu} for all numerical results.
	
	\begin{table*}[t]
		\centering
		\caption{Simulation parameters and values.}
		\label{tab:simu}
		\begin{tabular}{|p{5cm}|p{2.5cm}||p{5cm}|p{2.5cm}|}
			\hline
			\textbf{Parameters} & \textbf{Values} &\textbf{Parameters} & \textbf{Values} \\
			\hline 	
			Number of input ports & $M=2$ & Number of PAs & $N=10$ \\ \hline	
			Interval length of input ports & $L_{\text{IN}} = 5/4 \lambda_g$ & Interval length of PAs & $L_{\text{PA}}=1$ m \\	\hline 	
			Position of FU & $\mathbf{p}_{\text{FU}} = [5, 30]$ & Position of BU & $\mathbf{p}_{\text{BU}} = [-5, 20]$ \\
			\hline 	
			Maximum transmit power & $P_T=20$ dBm & Noise power & $N_0=-80$ dBm \\	\hline 	
			Initial value of penalty parameter & $\rho_0=0.1$ & Increase rate of $\rho$ & $c_{\rho}=1.02$ \\	\hline 	
			Maximum PA displacement & $\Delta=0.01$ m & Searching grid number of $d$ & $10^3$ \\	\hline 	
			Convergence tolerance & $10^{-3}$ m & Maximum iteration number & $I_{\max}=10^3$ \\	\hline 	
		\end{tabular}
	\end{table*}
	
	\subsection{Baseline Schemes}
	To verify the effectiveness of the proposed C-PASS architecture and the corresponding operating protocols, we compare with the following baseline schemes.
	\begin{itemize}
		\item \emph{End-fed PASS} (also referred to as conventional PASS): In this case, signals fed into $M$ input ports are directed towards the FP, radiating through $N$ FPAs to simultaneously serve both the FU and BU.
		
		\item \emph{Baseline 1} (also referred to as random transmit precoding): In this case, signals fed into $M$ input ports experience random precoding with the maximum power $P_T$. Specifically, we first randomly generate transmit precoding on fed signals, and then perform the remaining transmit and pinching beamforming to obtain the average sum rate over $200$ realizations.
		
		\item \emph{Baseline 2} (also referred to as uniform pinching beamforming): In this case, the positions of PA are uniformly distributed at the interval length of $L_{\text{PA}}$ without micro-adjustments, while the power radiation ratio of each PA is set as $\sqrt{1/N}$.  
		
	\end{itemize}

	\subsection{Convergence of Proposed Algorithms for PS, DS, and TS}
	\begin{figure}[t]
		\centering
		\includegraphics[width=1\linewidth]{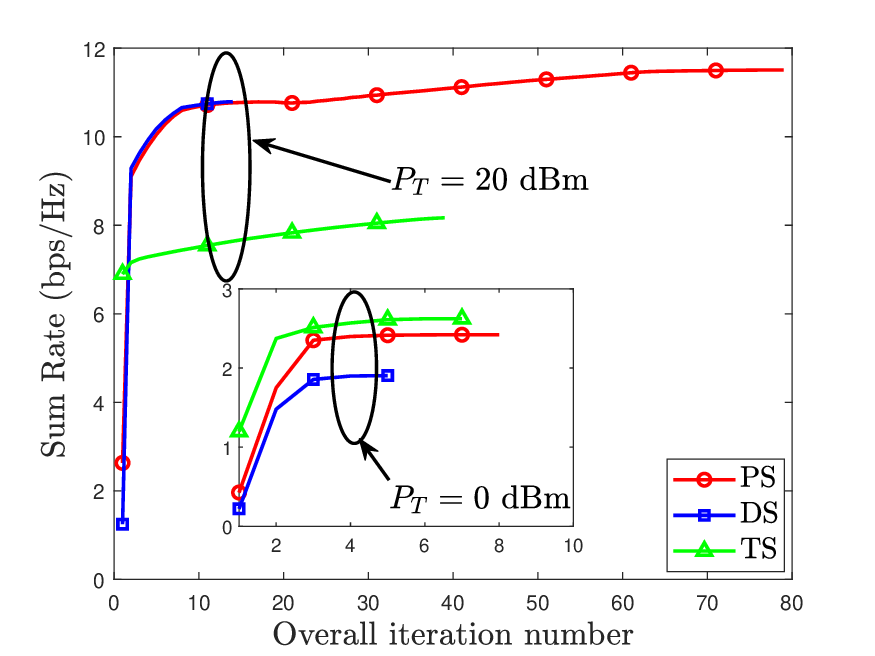}
		\caption{Coverage behaviors of the proposed algorithms.}
		\label{fig:simu_converage}
	\end{figure}
	
	In Fig.~\ref{fig:simu_converage}, we evaluate the convergence behavior of the proposed \emph{Algorithm~\ref{algorithm_PS_DS}} for the PS and DS protocols and \emph{Algorithm~\eqref{algorithm_TS}} for the TS protocol under the transmit powers of $P_T = 0$ and $20$ dBm. It can be observed that the sum rate of C-PASS for all considered protocols increases monotonically with the iteration number and rapidly approaches a stable value, which confirms the effectiveness and convergence of the proposed algorithms. Specifically, at $P_T = 0$ dBm, all the curves can converge within $10$ iterations. Fig.~\ref{fig:simu_converage} also shows that the PS protocol typically requires more iterations than DS, because the algorithm for PS involves the optimization of continuous power splitting ratios, whereas DS employs penalty-based methods for discrete boundaries. In addition, although TS attains a lower sum rate than PS and DS at $P_T=20$ dBm, it achieves the highest sum rate in the lower power regime at $P_T = 0$ dBm. This is because that, in the low-SNR regime, the C-PASS benefits primarily from the interference-free nature of TS for full transmit power on a single user per period. In contrast, in the high transmit power regime, the PS and DS protocols leverage flexible power splitting schemes to achieve superior spectral efficiency despite the presence of inter-user interference.

	\subsection{DoF Characterization of C-PASS}
	
	\begin{figure}[t]
		\centering
		\includegraphics[width=1\linewidth]{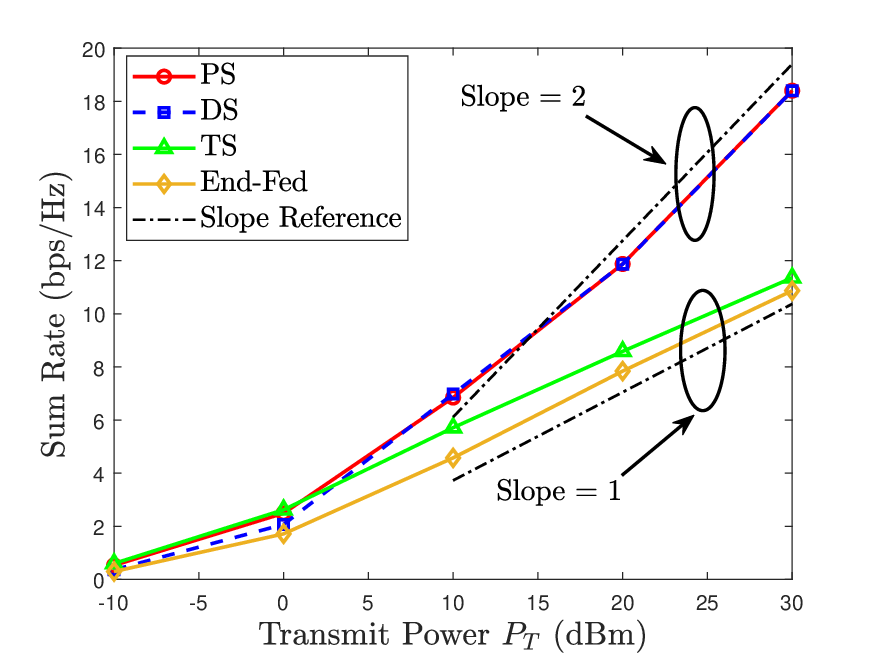}
		\caption{Sum-rate comparison under different protocols.}
		\label{fig:simu_DoF}
	\end{figure}
	
	Fig.~\ref{fig:simu_DoF} depicts the achievable sum rate versus transmit power for different protocols to evaluate the DoF performance. To quantify the slope of the curves in the high SNR regime, dotted reference lines with the slopes of $1$ and $2$ are provided. It can be observed that the sum-rate curves for the proposed PS and DS protocols asymptotically align with the reference line of $\text{slope} = 2$, whereas the curves for the TS protocol and the conventional end-fed architecture align with the reference line of $\text{slope} = 1$. This asymptotic behavior confirms that the PS and DS protocols achieve $\text{DoF}=2$, effectively doubling the spatial multiplexing capability compared to the conventional end-fed PASS with $\text{DoF}=1$. These results highlight the key advantage of the proposed C-PASS architecture in unlocking the potential for multi-stream transmissions. Furthermore, although the TS protocol exhibits a single DoF due to the time-orthogonal transmission, it still provides a clear improvement over the end-fed baseline. This performance gain arises from the additional directional flexibility enabled by the TS protocol, which allows the system to select the more favorable propagation direction for transmission.

	\subsection{Impact of Transmit Beamforming}
	\begin{figure}[t]
		\centering
		\includegraphics[width=1\linewidth]{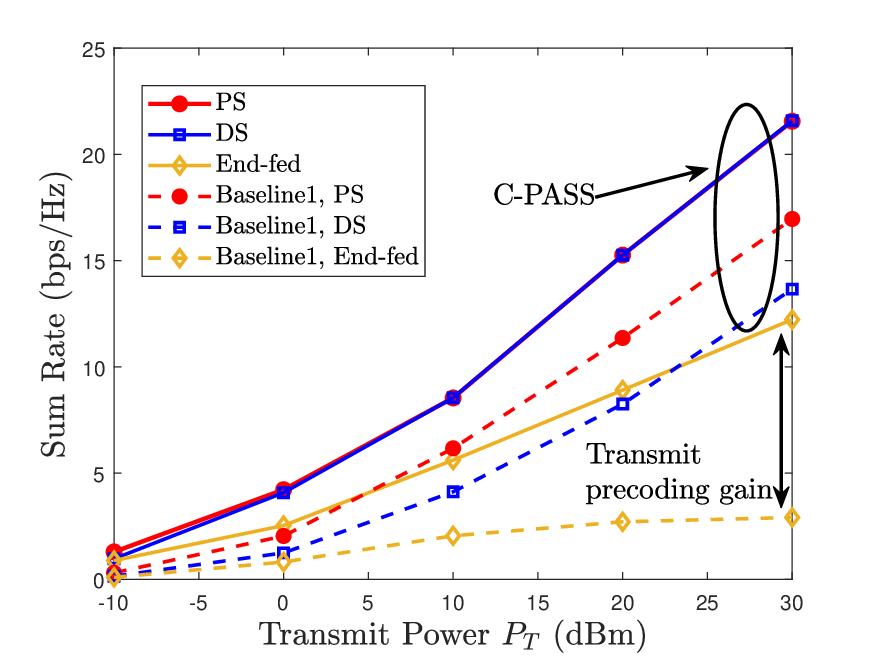}
		\caption{Sum-rate comparison with baseline 1, $M=6$. }
		\label{fig:simu_M6}
	\end{figure}
	Fig.~\ref{fig:simu_M6} illustrates the sum rate versus transmit power for the PS, DS, and conventional end-fed PASS, together with their corresponding baseline 1 schemes without transmit precoding. A clear performance improvement is observed for the proposed optimization schemes compared with their baseline 1, which highlights the importance of active transmit precoding before feeding into the waveguide. Moreover, it shows that the PS and DS protocols achieve comparable sum rates under the proposed joint optimization framework. However, in the scenario with random transmit precoding, the PS protocol significantly outperforms the DS. This behavior indicates that the optimization of the transmit precoder effectively mitigates the quantization loss inherent in the discrete DS constraints. Conversely, without active precoding alignment, the greater flexibility of the continuous power splitting ratios for the PS protocol provides a distinct performance advantage.
	
	Furthermore, despite the performance deterioration exhibited for the baselines 1 schemes on PS and DS, these curves maintain the same slope as their optimized counterparts in the high transmit power regime. This confirms that the DoF gain is inherent to the C-PASS architecture, independent of the specific transmit precoding design. Consequently, the PS and DS protocols with the baseline 1 scheme eventually outperform the end-fed architecture even when the latter employs optimal transmit precoding. This result highlights that the spatial multiplexing gain provided by the C-PASS architecture dominates in the high SNR regime.

	\begin{figure}[t]
		\centering
		\includegraphics[width=1\linewidth]{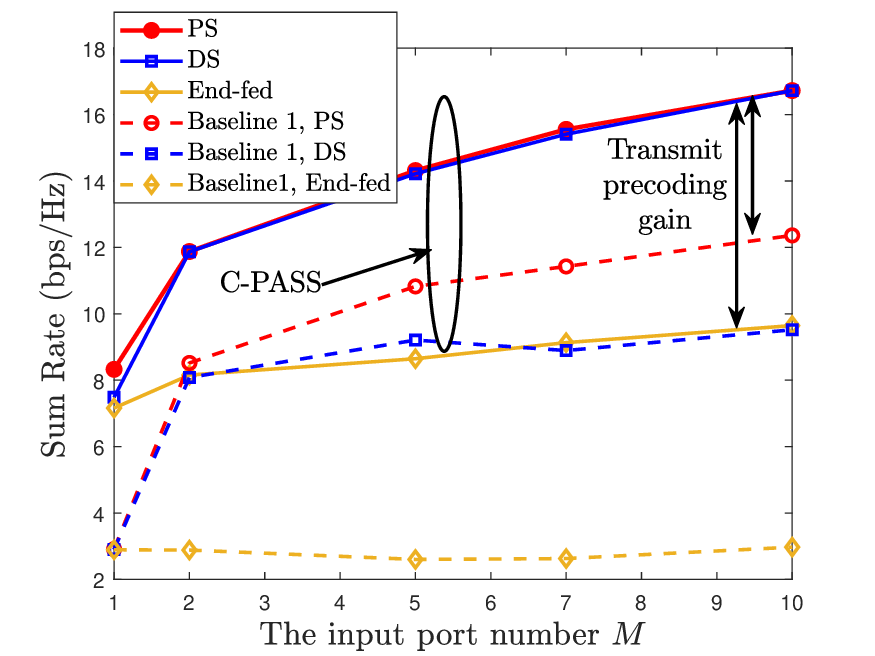}
		\caption{Sum rate versus $M$ under different protocols}
		\label{fig:simu_M_baseline1}
	\end{figure}
	
	In Fig.~\ref{fig:simu_M_baseline1}, we investigate the impact of the number of input ports $M$ on the achievable sum rate for the PS, DS, and conventional end-fed PASS. It can be observed that the sum rate for all schemes exhibits a monotonic increase with $M$. This trend is expected as a larger array of input ports enables a higher transmit beamforming gain. More particularly, at $M=1$, the sum-rate performances of PS, DS, and end-fed PASS are nearly identical. This is because that the single input port restricts the achievable DoF to one for all schemes. As $M$ increases from $1$ to $2$, a substantial performance improvement is observed for PS and DS as well as their baseline 1, which reflects that the C-PASS architecture can effectively unlock its potential of the doubled DoF. Furthermore, the performance advantage of C-PASS over the conventional end-fed PASS becomes increasingly prominent as $M$ grows. This is because that the additional input ports allow the more spatial multiplexing capability of C-PASS to be better exploited. Moreover, a larger performance improvement is also observed between PS and DS compared with their baseline 1 schemes, indicating that when transmit precoding is random, the continuous power splitting ratios for PS become more beneficial, particularly for larger $M$.

	\subsection{Impact of Pinching Beamforming}
\begin{figure}[t]
		\centering
		\includegraphics[width=1\linewidth]{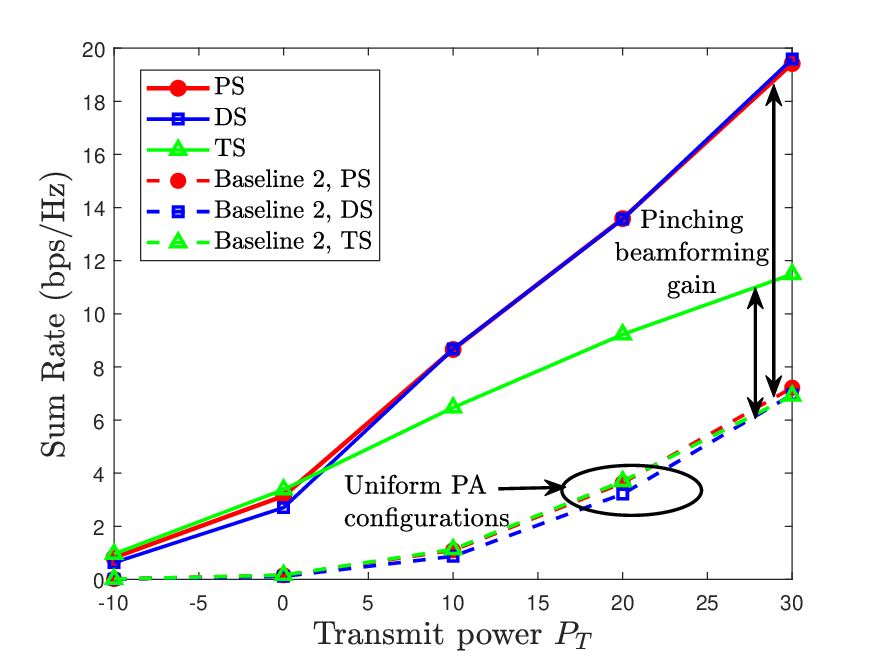}
		\caption{Sum-rate comparison with baseline 2, $N=20$.}
		\label{fig:simu_N20}
	\end{figure}

	In Fig.~\ref{fig:simu_N20}, we plot the sum rate versus transmit power for the PS, DS, and TS protocols and their corresponding baseline 2 scheme with uniform PA configurations, under $N=20$. A substantial performance gain is observed between the proposed optimization framework and its uniform baselines, validating the critical importance of jointly optimizing the PA positions and power radiation coefficients. Specifically, the proposed algorithm for the PS and DS protocols yields a sum-rate gain of approximately $4.5$ dB over their baseline 2 schemes, underscoring the advantages of the pinching beamforming design.
	
	Compared with the results in Fig.~\ref{fig:simu_DoF}, the sum rate improvements of PS and DS compared with TS become more evident at the larger PA number of $N=20$. This is because that PS and DS can simultaneously exploit the radiated signals from both FP and BP directions, effectively utilizing $2N$ PAs, whereas TS activates only one propagation direction within each time period and therefore leverages only $N$ PAs. As a result, the array gains from PAs offered by PS and DS are more effectively translated into sum-rate improvements when the number of PAs is large. Furthermore, it can be observed that the baselines 2 for all three protocols exhibit nearly identical performance across the entire transmit power range from $-10$ dBm to $30$ dBm. This is because that, for the baseline 2 scheme, a large portion of PAs are located at long distances from the users, leading to severe path loss attenuation. Consequently, the received signal strength is mainly governed by the overall array gain rather than the spatial multiplexing gain.

	\begin{figure}[t]
		\centering
		\includegraphics[width=1\linewidth]{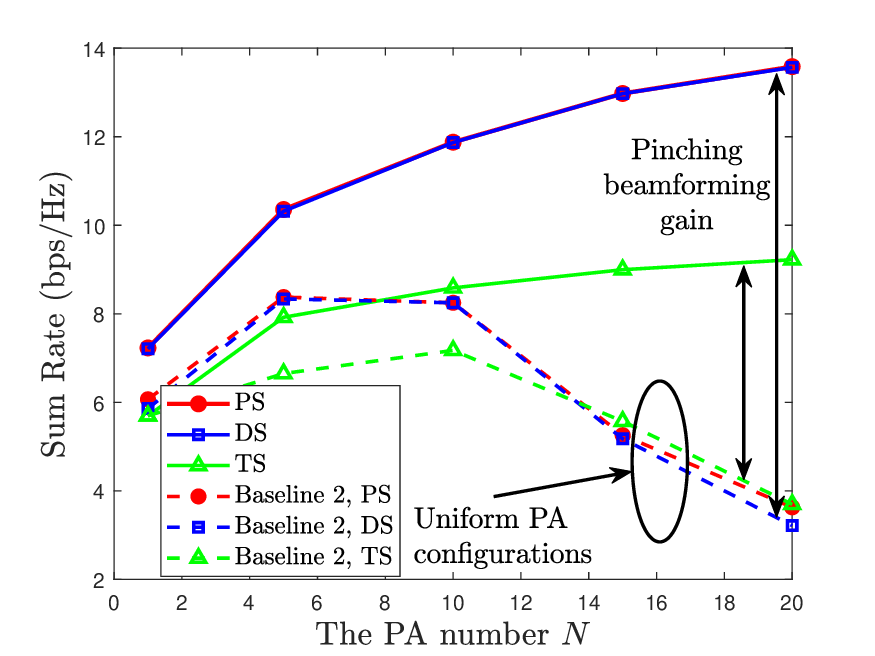}
		\caption{Sum rate versus $N$ under different protocols.}
		\label{fig:simu_N_baseline2}
	\end{figure}
	
	Fig.~\ref{fig:simu_N_baseline2} investigates the impact of the number of PAs $N$ on the achievable sum rate for the PS, DS, and TS protocols. It shows that the sum rate for the proposed optimization framework exhibits a monotonic increase with $N$ across all three protocols. This trend confirms that the proposed pinching beamforming design can effectively exploit the array gain provided by a larger number of PAs to enhance signal focusing. In contrast, the sum-rate performance for their baseline 2 schemes exhibits a concave trend, initially increasing and eventually degrading as $N$ becomes large. This non-monotonic behavior arises because the uniform configuration of PAs fails to adaptively concentrate the radiated power. Consequently, the performance improvement of the proposed optimization framework compared with their baseline 2 schemes becomes increasingly pronounced at larger $N$. Specifically, the proposed pinching beamforming yields a substantial sum-rate enhancement of approximately $4.0$ dB for the TS protocol, while the gain for the PS protocol reaches as high as $5.7$ dB, further validating the significance of precise PA placement and power radiation optimization.
	
	\section{Conclusion}\label{sec:conclusion}
	The C-PASS designs for multi-user wireless communications were investigated. Based on the basic signal model at each input port, three practical operating protocols for C-PASSs were proposed, and their respective optimization variables and advantages were discussed. For each of these operating protocols, the joint transmit and pinching beamforming optimization problem was formulated to maximize the system sum rate. More particularly, the resulting highly coupled non-convex problems were efficiently solved by WMMSE reformulation and alternating optimization algorithms. Then, the proposed algorithm was extended with the penalty-based optimization for DS. For the TS protocol, the optimization of time allocation ratios was solved in closed form. Numerical results showed that C-PASS can significantly enhance the system sum rate compared with conventional PASS. Furthermore, the obtained results also revealed that the transmit beamforming gain for the PS and DS is increasing with $P_T$ and $M$, while the pinching beamforming gain increases with $P_T$ and $N$. 
	
	These results confirm the effectiveness of employing C-PASSs for improving the communication performance of wireless networks, which motivates related future research on C-PASSs. In particular, the C-PASS for the proposed protocols provides significant design flexibility compared with the conventional PASS. By optimizing the power splitting ratios for the PS and DS protocols or the time allocation ratios for the TS protocol, the system fully exploits the joint transmit and pinching beamforming gains and leverages the additional DoF for multi-stream transmission. Furthermore, these properties offer promising directions for future investigation into uplink and downlink communications, where C-PASS is expected to significantly improve channel estimation, data detection, and transmission reliability, especially in multi-user scenarios.

	\bibliographystyle{IEEEtran}
	\bibliography{reference/mybib}

\end{document}